\documentclass[seceq,preprint]{ptptex}

\def\nn{\nonumber\\}

\def\B{{\rm B}}
\def\QB{Q_{\rm B}}
\def\half{{1\over2}}
\newcommand{\bra}[1]{\left\langle#1\right\vert}
\newcommand{\ket}[1]{\left\vert#1\right\rangle}

\def\preBRST{\delta_\B}

\def\YY{Y\bar Y}

\def\fket#1{\bigl|#1\bigr\rangle}
\def\fbra#1{\bigl\langle#1\bigr|}
\def\fbraket#1#2{\bigl\langle#1\bigr|#2\bigr\rangle}

\def\Gpsim{G_{-\half}\ket{\psi_{+}}+G_{\half}\ket{\psi_{-}}}
\def\Gphip{G_{\half}\ket{\phi_{+}}+G_{-\half}\ket{\phi_{-}}} 
\def\Gphim{G_{-\half}\ket{\phi_{+}}+G_{\half}\ket{\phi_{-}}}

\newcommand{\braket}[2]{\left<#1|#2\right>}
\newcommand{\nket}[1]{\!\left.\vphantom{\mathstrut}\right|\!#1\!\left>\vphantom{\mathstrut}\right.\!}
\newcommand{\nbra}[1]{\!\left<\vphantom{\mathstrut}\right.\!#1\!\left|\vphantom{\mathstrut}\right.\!}
\newcommand{\nbraket}[2]{\!\left<\vphantom{\mathstrut}\right.\!#1|#2\!\left>\vphantom{\mathstrut}\right.\!}
\newcommand{\dket}[1]{\vert#1\rangle\!\rangle}
\newcommand{\dbra}[1]{\langle\!\langle#1\vert}

\newcommand{\Exp}[1]{\,e^{#1}}
\def \ii {i}

\def \dd {\partial}
\def \tQB {{\tilde Q}_{\rm B}}

\def \delB {\delta_{\rm B}}

\def \Ramond {\Psi}
\def \NS {\Phi}
\def \PRamond {{\hat \Ramond}} 
\def \bPRamond {{\bar \PRamond}}
\def \PPRamond  {\PRamond_{\Vert}{}}
\def \bPPRamond  {{\bar \PRamond}_{\Vert}{}}
\def \PNL {\hat B}
\def \PLambda {{\hat \Lambda}}

\def \PiR {\Pi_{\rm R}}

\def \Xm {X_{-\frac12}}
\def \Xp {X_{\frac12}}

\def \Prj {\mathcal P}

\def \VNNN {\bra{V_{\text{NS-NS-NS}}}}
\def \VNRR {\bra{V_{\text{NS-R-R}}}}
\def \VI {\bra{V_I}}

\notypesetlogo                       
\preprintnumber[3cm]{
YITP-11-98\\ November, 2011}

\title{
Gauge Fixing of Modified Cubic \\
Open Superstring Field Theory
}

\author{
Maiko \textsc{Kohriki},\footnote{%
E-mail:\  {kohriki@yukawa.kyoto-u.ac.jp}} %
Taichiro \textsc{Kugo}\footnote{%
E-mail:\  {kugo@yukawa.kyoto-u.ac.jp}} %
and %
Hiroshi \textsc{Kunitomo}\footnote{%
E-mail:\  {kunitomo@yukawa.kyoto-u.ac.jp}}  
}

\inst{
Yukawa Institute for Theoretical Physics, Kyoto University, \\
Kyoto 606-8502, Japan
}

\abst{
The gauge-fixing problem in the modified cubic open superstring field theory
is discussed in detail for both the Ramond and Neveu-Schwarz (NS) sectors
in the Batalin-Vilkovisky (BV) framework. We prove
for the first time that the same form of action as the classical
gauge-invariant one with the ghost-number constraint on the string
field relaxed, gives the master action satisfying the BV master equation.
This is achieved by identifying independent component fields
by analyzing the kernel structure
of the inverse picture-changing operator. The explicit
gauge-fixing conditions for the component fields are discussed.
In a kind of $b_0=0$ gauge, we explicitly obtain an NS propagator
that has poles at zeros of the Virasoro operator $L_0$.
}

\begin{document}

\maketitle

\section{Introduction}

The covariant open superstring field theory (super-SFT) was first 
constructed by Witten\cite{Witten:1986qs} 
in the form of Chern-Simons three-form action 
similarly to his bosonic string field theory.\cite{Witten:1985cc}
He wrote the action using string fields in the natural picture, 
that is, the Neveu-Schwarz 
(NS) string of picture number $-1$ and the Ramond (R) string of picture 
number $-1/2$. The action had a picture-changing operator $X(z)$ located
at the midpoint $z=i$ in the cubic NS interaction term. Later, however, 
it was pointed out by Wendt\cite{Wendt:1987zh} 
that the theory suffers from a severe 
divergence problem caused by colliding picture-changing operators in 
the NS four-string amplitude.
This led to the violation of the associativity, and hence, the 
gauge invariance of the theory.

To circumvent this above problem, two approaches were proposed. One is the 
so-called modified cubic super-SFT, which has the same form of action as 
Witten's but is based on the $0$-picture NS string field, and was 
proposed by Preitschopf, Thorn, and Yost\cite{Preitschopf:1989fc} (PTY) 
and Arefeva, Medvedev, and Zubarev\cite{Arefeva:1989cp} (AMZ), 
independently. Another is Berkovits' 
super-SFT,\cite{Berkovits:1995ab} 
which has the WZW-type nonpolynomial action for the NS sector. 
The latter approach is interesting in the sense that there is no need for 
a picture-changing operator. 
However, in this WZW type approach, 
the construction of the action for the R sector is difficult and has never been performed in a satisfactory manner.\cite{Michishita:2004by}

The modified cubic super-SFT approach, on the other hand, simply gives the R 
sector as a cubic action and is beset with no problem. 
Therefore, we study the modified cubic super-SFT in this paper. 
The above-mentioned authors who proposed the modified cubic super-SFT discussed the perturbation theory 
in their papers so that they must have discussed the gauge-fixing problem. 
However, strangely enough, they did not show that their gauge-fixed 
action is really BRST-invariant. 
First, they {\em tacitly} assumed that 
the gauge-fixed action, or even the Batalin-Vilkovisky (BV) master 
action for fields and antifields, takes the {\em same form} as the 
original gauge-invariant action with the understanding that the field 
ghost number constraint is relaxed. 
However, to the best of our knowledge, no one has ever shown that the `BV action' thus 
obtained does really satisfy the BV master equation and hence that the 
gauge-fixed action obtained by fixing the gauge by setting the suitable 
set of antifields equal to zero is really BRST-invariant. 

The invariance of the `BV action' under the BRST transformation 
$\delta_\B\Phi= \QB\Phi+\Phi*\Phi$ is trivial by construction but the BV master equation is not because of the presence of inverse 
picture-changing operators that have nontrivial kernels and thus are non-invertible. 
The presence of nontrivial kernels of inverse picture 
changing operators introduces an additional gauge invariance other than the 
conventional one. 
This is important for the very consistency of the R sector 
in Witten's original super-SFT. This is also the same in this modified 
cubic super-SFT with the same R sector. 

The NS sector is more difficult in which a double-step inverse 
picture-changing operator 
is present in the modified cubic super-SFT case. The kernel 
becomes much larger than that in the R case and the construction of 
the BRST-invariant projection operator becomes much more complicated 
accordingly. 
Because of this complication in the projection operator,\footnote{Many trials have also been carried out to find simpler BRST-invariant projection operators or even other possibilities of the double-step inverse picture-changing operator for which projection operators become simpler.\cite{Kohriki:2011zz}}
no explicit analysis of component fields of the NS field reduced into the projected space has been carried out; 
hence, no explicit proof of the BV master equation has been 
given. The validity of the gauge-fixing procedure could be judged only 
with it. 

The purpose of the present paper is to discuss the gauge fixing of the 
modified cubic super-SFT, particularly to give an explicit proof that 
the same form of action as the original gauge-invariant super-SFT action 
with the ghost number constraint relaxed satisfies the BV master equation.
This can be achieved by adopting a simple component field form of the 
reduced string field suggested directly from the form of the inverse 
picture-changing operator. 
This reduced form is not invariant under the BRST 
transformation, but it can be modified so as to keep the form.

This paper is organized as follows. We first recall the gauge-invariant action of the modified cubic open super-SFT in 
\S\ref{sec:gaugeinv-action_MCSSFT}. 
Then, in order to understand the nature of the problem, 
we review in \S\ref{sec:gaugefix-R} 
the gauge-fixing procedure for the R sector, 
which was beautifully performed by us and Masaki Murata. 
This procedure was reported in the string field theory workshops held at APCTP in 
2009\cite{Kugo:2009APCTP} and, at YITP in 2010,\cite{Kohriki:2011zz} 
but the full report has never been published. 
The kernel problem in the NS sector is much more complicated than that
in the R sector, but we give a solution to this problem in 
\S\ref{sec:NS_reducedform-BV}, as 
mentioned above, by defining the simple reduced 
form of the NS string field for which the double-step picture-changing 
operator $\YY$ gives a nondegenerate cross-diagonal metric. 
We show there that it implies that the action satisfies the BV master equation. 
In \S\ref{sec:gf-component-NS}, 
we discuss the gauge fixing for the NS sector by explicitly 
examining the component fields in the reduced NS field.
We propose a set of gauge-fixing conditions that gives 
a BRST-invariant consistent gauge-fixed action. 
However, since it is difficult to find a propagator in a closed form in that gauge, we propose another 
gauge in \S\ref{sec:NSgf_nonlocal-gc} using a nonlocal projection operator.
It can give a propagator explicitly, but the 
explicit component expression becomes much less clear. 
The gauge is of the $b_0=0$ type and the propagator has poles at $L_0=0$.
A note on the the gauge fixing for the total interacting  system is 
given in \S\ref{sec:gf_interacting}.
The final section, \S\ref{sec:discussion}, is devoted to discussions. 
We add three appendices.
The explicit expressions for the BRST operators and BRST 
transformations are given in Appendix \ref{app:BRSTop-trf}.
The definitions and properties of the picture-changing operators 
are summarized in Appendix \ref{app:PCO}. 
We give the component expressions for the decomposition by nonlocal 
projection operators for the R sector in Appendix \ref{app:comp-nonlocal}.

\section{Gauge-invariant action of modified cubic open super-SFT}
\label{sec:gaugeinv-action_MCSSFT}

The modified cubic open super-SFT action based 
on the 0-picture 
NS string $\Phi$ and $(-1/2)$-picture R string 
$\Psi$ was proposed by PTY \cite{Preitschopf:1989fc} and AMZ, \cite{Arefeva:1989cp} independently, and is given by
\begin{align}
S 
&= \half\int\YY \Phi * \QB \Phi  + {1\over3} \int\YY \Phi * \Phi * \Phi  
 + \half\int Y \Psi * \QB \Psi+  \int Y \Phi * \Psi* \Psi
\nn
&= \half\bra{\Phi} \YY \QB \ket{\Phi}  
 + {1\over3} \VNNN  \YY \ket{\Phi} \ket{\Phi} \ket{\Phi}  
\nn
&\hspace{18.75ex}
 - \half \nbra{\bar \Psi} Y \QB \ket{\Psi} 
 + \VNRR Y \ket{\Phi} \ket{\Psi} \ket{\Psi}.
\label{eq:action_NS-R}
\end{align}
In the classical action the string fields $\NS$ and $\Ramond$ are constrained to possess ghost number one.\footnote{This corresponds to the requirement that the coefficient fields carry ghost number zero.} 
The symbol $*$ denotes the conventional Witten's string product based on the midpoint interaction.
The operators $Y=Y(\ii)$ and $\bar Y=Y(-\ii)$ are the inverse picture-changing operators inserted at the string midpoint $z=\ii$ and its mirror point $z=-\ii$.\footnote{Thus we adopt the nonchiral operator as the double-step inverse picture-changing operator following PTY. The conventions of the picture-changing operators are summarized in Appendix \ref{app:PCO}.}
Note that $\nbra{\bar \Ramond}$ is the Dirac conjugate of $\ket{\Ramond}$:
\begin{equation}
\nbra{\bar \Ramond} = \nbra{\Ramond} \psi^0_0,
\qquad  \nbra{\Ramond} = (\ket{\Ramond})^\dagger,
\end{equation}
where $\psi^\mu_0$ is the zero mode of the matter operator $\psi^\mu(z)$ in the R sector.

This system is invariant under the gauge transformations
\begin{subequations} \label{eq:gaugetrfQB}
\begin{eqnarray}
\delta\Phi &=& \QB \lambda+ \Phi * \lambda- \lambda*\Phi + \bar X(\Psi*\chi-\chi*\Psi), 
\label{eq:gaugetransformNS}
\\ 
\delta\Psi&=& \QB \chi+ \Phi * \chi- \chi* \Phi + \Psi* \lambda-\lambda*\Psi, 
\label{eq:gaugetransformR}
\end{eqnarray}
\end{subequations}
where $\lambda$ ($\chi$) is the NS (R) string transformation parameter with ghost number zero and $\bar X = X(-\ii)$ is the picture-changing operator at the mirror point.

Furthermore, the marked difference from the bosonic string case 
is that the `measures' $\YY= Y(i)Y(-i)$ in the NS sector
and $Y=Y(\ii)$ in the R sector have nontrivial kernels since 
\begin{equation} \label{eq:Y-kernel}
c(z)Y(z) = \gamma^2(z)Y(z) =0,
\end{equation}
as is clear from the expression
\begin{equation}
Y(z) = c(z)\delta'(\gamma(z)).
\end{equation}
Consequently, the kinetic terms of the action \eqref{eq:action_NS-R} are also invariant under additional gauge transformations:\cite{Kugo:1988mf, Arefeva:1988nn}
\begin{alignat}{2}
\delta \NS &= c(\ii)\lambda_c,
&\qquad 
\delta \NS &= \gamma^2(\ii)\lambda_\gamma,
\\
\delta \NS &= c(-\ii)\lambda_c', 
&\qquad 
\delta \NS &= \gamma^2(-\ii)\lambda_\gamma'.
\\
\delta \Ramond &= c(\ii)\chi_c,
&\qquad 
\delta \Ramond &= \gamma^2(\ii)\chi_\gamma.
\label{eq:YkernelR}
\end{alignat}
This is actually the symmetry of the total action \eqref{eq:action_NS-R} including the interaction terms because the ghost fields $\gamma(z)$ and $c(z)$ vanish at the interaction point:
\begin{equation} \label{eq:vprop}
\VI c(\pm\ii) = \VI \gamma(\pm\ii) =0,
\qquad 
(I=\text{NS-NS-NS or NS-R-R})
\end{equation}
where $c(\pm\ii)$ or $\gamma(\pm\ii)$ can be the ghost coordinate of any of the three strings.

Since the additional gauge invariances are different between the R and NS cases, we discuss the two sectors separately, initially neglecting the Yukawa coupling term $\NS\Ramond\Ramond$.

\section{Gauge fixing of Ramond sector}
\label{sec:gaugefix-R}

In this section, we discuss the gauge fixing of the gauge symmetries \eqref{eq:gaugetransformR} and \eqref{eq:YkernelR} for the R string. 
This can be performed beautifully. 
We will provide a prototype solution to the problem.

We first fix the additional gauge invariance \eqref{eq:YkernelR} using the BRST-invariant projection operator $\Prj_Y$ following Ref.~\citen{Kugo:1988mf}:
\begin{equation}
\Prj_Y \equiv X_0 Y,  \qquad 
X_0 = [\QB, \Theta(\beta_0)] = \delta(\beta_0) F_0 - b_0\delta'(\beta_0),
\end{equation}
where $F_0 = [\QB,\beta_0]$.
The choice of $X_0$ is not unique and has the freedom of changing the gauge for new gauge symmetries. 
Here, $X_0$ is the `mode version' of the picture-changing operator satisfying the properties \eqref{eq:XYX_Ramond} in Appendix \ref{app:PCO}.

The R string field $\Psi$ can be generally expanded in the zero modes of the ghost $c_0$ and the superghost $\gamma_0$ as
\begin{align}
\ket{\Psi}
&= \sum_{n=0}^\infty (\gamma_0)^n \ket{0}_{\beta} \otimes 
    \Big( \ket{\downarrow} \otimes \ket{\phi_n} 
	+ c_0 \ket{\downarrow} \otimes \ket{\psi_n}\Big)
\\
&\equiv \sum_{n=0}^\infty (\gamma_0)^n 
    \Big( \dket{\phi_n} + c_0 \dket{\psi_n}\Big),
\end{align}
where $\ket{\downarrow}$ and $\ket{0}_{\beta}$ are, respectively, the vacua of the reparametrization ghost zero mode $b_0$ and the superghost zero mode $\beta_0$ defined by
\begin{alignat}{2}
b_0\ket{\downarrow}&=0, 
&\qquad \bra{\downarrow}c_0\ket{\downarrow}&=1,
\\
\beta_0 \ket{0}_{\beta} &= 0, 
&\qquad 
{\vphantom{\ket{0}}}_{\beta}\!\bra{0} \delta(\gamma_0) \ket{0}_{\beta} &=1.
\end{alignat}
The states $\ket{\phi_n}$ and $\ket{\psi_n}$ are the states in the sector other than the (super)ghost zero modes.
For brevity, we write $\ket{0}_{\beta} \otimes \ket{\downarrow} \otimes \ket{\varphi}$ as $\dket{\varphi}$, which satisfies
\begin{equation} \label{eq:dket-ket}
\dbra{\varphi_1} c_0 \delta(\gamma_0) \dket{\varphi_2} = \braket{\varphi_1}{\varphi_2}.
\end{equation}

The projected field has been shown\cite{Kugo:1988mf} to take the same form as the constrained string field proposed in Ref.~\citen{Kazama:1986cy}:
\begin{equation}
\nket{\PRamond} \equiv
\Prj_Y\Psi =  
 \dket{\phi} - \left(\gamma_0+c_0 F\right) \dket{\psi},
\end{equation}
where $F =  F_0 + 2b_0\gamma_0$ is the Ramond-Dirac operator \eqref{eq:mode-F_R} with the ghost zero modes $b_0, \gamma_0$ removed  from $F_0$. 
Henceforth, the hatted field denotes the field whose kernel degrees of freedom are projected out by the BRST-invariant projection operator $\Prj_Y$.
The Ramond kinetic term can be rewritten with the projected field as 
\begin{align}
S[\PRamond] 
&= - \frac12 \nbra{\bar \Ramond} \Prj_Y^T Y \QB \Prj_Y \ket{\Ramond}
 = - \frac12 \bra{\Ramond} \psi^0_0 Y X_0 Y \QB X_0 Y \ket{\Ramond}
\nn
&= - \frac12 \bra{\Ramond} Y X_0^\dagger \psi^0_0 Y \QB X_0 Y \ket{\Ramond}
 = - \frac12 \bra{X_0 Y \Ramond} \psi^0_0 Y \QB X_0 Y \ket{\Ramond}
\nn
&= - \frac12 \nbra{\PRamond} \psi^0_0 Y \QB \nket{\PRamond}
 = - \frac12 \nbra{\bar \PRamond} Y \QB \nket{\PRamond}.
\end{align}
Here we have used $\{Y,\psi^0_0\}=0$ and Eq.~\eqref{eq:X_0dagger}.

Next, for the conventional $\QB$ gauge invariance, 
we impose the gauge-fixing condition 
\begin{equation}
\ket{\psi}=0.
\end{equation}
Then the projected field $\PRamond$ satisfies 
the following two conditions simultaneously:
\begin{equation} \label{def:Ramond-Siegel_gauge}
\beta_0 \nket{\PRamond} = 0, 
\quad \text{and} \quad 
b_0 \nket{\PRamond} = 0.
\end{equation}
Let us call this the Ramond-Siegel gauge. 
We introduce the projection operator into the Ramond-Siegel gauge subspace:
\begin{equation}
\Prj_G \equiv b_0c_0 \delta(\beta_0) \delta(\gamma_0),  \qquad 
\nket{\PPRamond} \equiv \Prj_G \nket{\PRamond} = 
\dket{\phi}.
\end{equation}

\subsection{Iterative gauge-fixing procedure}

For the free action, the gauge fixing can be carried out iteratively just as in the bosonic case.\cite{Terao:1986ex}\tocite{Kiermaier:2007jg}
Let us start from the gauge-invariant free action:
\begin{equation} \label{eq:S0-Ramond}
S_0
 = - \frac12 \fbra{\bPRamond_{(1)}} Y_0 \QB \fket{\PRamond_{(1)}}
 = - {1\over2} \nbra{{\bar \Ramond}_{(1)}} \Prj_Y^T Y_0 \QB \Prj_Y \nket{\Ramond_{(1)}},
\end{equation}
where $\nket{\Ramond_{(1)}}$ is the classical R string field with ghost number one.
Note that the inverse picture-changing operator $Y$ can be rewritten as $Y_0 = c_0\delta'(\gamma_0)$ between the projection operators using Eqs.~\eqref{eq:XYX_Ramond}:
\begin{equation}
\Prj_Y^\dagger Y \Prj_Y =\Prj_Y^\dagger Y_0 \Prj_Y.
\end{equation}
The above action \eqref{eq:S0-Ramond} is invariant under the gauge transformation with the gauge parameter of ghost number zero:
\begin{equation}
\delta\nket{\PRamond_{(1)}} = \QB \nket{\PLambda_{(0)}}. 
\label{eq:gaugetrfS0-R}
\end{equation}
We can rewrite the gauge transformation \eqref{eq:gaugetrfS0-R} as the BRST transformation by introducing the Fadeev-Popov (FP) ghost field $\nket{\PRamond_{(0)}}$ and 
replacing the parameter $\nket{\PLambda_{(0)}}$ with this field as
\begin{equation}
\delta_\B \nket{\PRamond_{(1)}} = \QB \nket{\PRamond_{(0)}}.
\end{equation}
It is convenient to introduce the FP antighost field $\nket{\PRamond_{(2)}}$ and the Nakanishi-Lautrup (NL) field $\nket{\PNL_{(2)}}$ with the BRST transformations
\begin{equation}
\delta_\B \nket{\PRamond_{(2)}} = \nket{\PNL_{(2)}}, 
\qquad 
\delB \nket{\PNL_{(2)}} = 0.
\end{equation}
Taking the Ramond-Siegel gauge condition 
\begin{equation}
(1-\Prj_G) \nket{\PRamond_{(1)}} = 0,
\end{equation}
we add the gauge-fixing and the FP ghost terms to the gauge-invariant action \eqref{eq:S0-Ramond}:
\begin{align}
S_1 
&= S_0 
 - \delB \left(\fbra{\bPRamond_{(2)}} Y_0 (1-\Prj_G) \fket{\PRamond_{(1)}} \right) 
\nn
&= - {1\over2} \fbra{\bPRamond_{(1)}} Y_0 \QB \fket{\PRamond_{(1)}} 
 - \fbra{{\bar \PNL}_{(2)}} Y_0 (1-\Prj_G) \fket{\PRamond_{(1)}} 
 - \fbra{\bPRamond_{(2)}} \Prj_G^T Y_0 \QB \fket{\PRamond_{(0)}}.
\label{eq:S1-Ramond}
\end{align}
Here, we have used the equality
\begin{equation}
Y_0 (1-\Prj_G) = \Prj_G^T Y_0,
\end{equation}
which can be easily confirmed by the following direct calculations:
\begin{align}
Y_0 (1-\Prj_G)
&= Y_0 - c_0 [\delta(\gamma_0), \beta_0] b_0 c_0 \delta(\beta_0) \delta(\gamma_0) 
\nn
&= Y_0 + c_0 \beta_0 \delta(\gamma_0) 
 = c_0 \delta(\gamma_0) \beta_0, 
\\
\Prj_G^T Y_0 
&= c_0 b_0 \delta(\gamma_0) \delta(\beta_0) c_0 [\delta(\gamma_0), \beta_0]
 = c_0 \delta(\gamma_0) \beta_0. \label{eq:PgdaggerY_0}
\end{align}
This action $S_1$ is still invariant under the gauge transformation
\begin{equation}
\delta \nket{\PRamond_{(0)}} = \QB \nket{\PLambda_{(-1)}}.
\end{equation}

Repeating the same procedure successively, we can obtain the totally gauge-fixed action as follows.
First, we introduce a series of FP ghost fields $\nket{\PRamond_{(-g)}}$, FP antighost fields $\nket{\PRamond_{(2+g)}}$, and NL fields $\nket{\PNL_{(2+g)}}$ for $g\geq0$ with the BRST transformations
\begin{alignat}{2}
&\delta_\B \nket{\PRamond_{(1-g)}} = \QB \nket{\PRamond_{(-g)}},
&\qquad
\delta_\B \nket{\PRamond_{(2 + g)}} &= \nket{\PNL_{(2 + g)}},
\nn
&\delB \nket{\PNL_{(2 + g)}} = 0,
&\qquad 
\text{for} \quad g&\geq 0.
\label{eq:BRSTtrf_FP-aFP-NL}
\end{alignat}
Imposing the Ramond-Siegel gauge condition $(1-\Prj_G) \nket{\PRamond_{(1-g)}} = 0$ for the FP ghosts and the original field, we obtain
\begin{align}
S &= - {1\over2} \fbra{\bPRamond_{(1)}} Y_0 \QB \fket{\PRamond_{(1)}} 
 - \sum_{g=0}^\infty \delta_\B \left(\fbra{\bPRamond_{(2+g)}} Y_0 (1-\Prj_G) \fket{\PRamond_{(1-g)}} \right),
\nn
&= - {1\over2} \fbra{\bPRamond_{(1)}} Y_0 \QB \fket{\PRamond_{(1)}} 
\nn
&\quad\,
 -  \sum_{g=0}^\infty
   \left(
    \fbra{{\bar \PNL}_{(2+g)}} Y_0 (1-\Prj_G) \fket{\PRamond_{(1-g)}} 
    + \fbra{\bPRamond_{(2+g)}} \Prj_G^T Y_0 \QB \fket{\PRamond_{(-g)}}
   \right).
\label{eq:Sn-Ramond}
\end{align}
The NL fields $\nket{\PNL_{(2+g)}}$ serve as gauge-fixing multipliers corresponding to the Ramond-Siegel gauge conditions for the FP ghost fields $\nket{\PRamond_{(-g)}}$ and the original field $\nket{\PRamond_{(1)}}$:
\begin{equation}
(1-\Prj_G) \nket{\PRamond_{(1-g)}} = 0 \quad \text{for}\quad g\geq0.
\label{gauge-conditionR}
\end{equation}
In addition, we can see from the action \eqref{eq:Sn-Ramond} that antighost fields are projected by $\Prj_G$ and satisfy the Ramond-Siegel gauge condition automatically:
\begin{equation}
(1-\Prj_G) \Prj_G \nket{{\PRamond}_{(2+g)}} = 0 \quad \text{for}\quad g\geq0.
\end{equation}
Integrating out the NL fields $\nket{\PNL_{(2+g)}}$, therefore,
we can replace all the string fields $\nket{\PRamond_{(g)}}$ by the gauge-fixed fields $\nket{\PPRamond_{(g)}}=\Prj_G\nket{\PRamond_{(g)}}$ and obtain 
\begin{align}
S &= - {1\over2} \fbra{\bPPRamond_{(1)}} Y_0 \QB \fket{\PPRamond_{(1)}} 
 - \sum_{g=0}^\infty \fbra{\bPPRamond_{(2+g)}} Y_0 \QB \fket{\PPRamond_{(-g)}},
\nn
&= - {1\over2} \fbra{\bPPRamond} Y_0\QB \fket{\PPRamond},
\label{eq:Sgaugefixed-Ramond}
\end{align}
where 
\begin{equation}
\nket{\PPRamond} \equiv \sum_{g=-\infty}^\infty \nket{\PPRamond_{(g)}}.
\end{equation}
This action \eqref{eq:Sgaugefixed-Ramond} has the same form as the original action, but with the ghost number constraint relaxed, and is subject to the gauge condition
\begin{equation} \label{eq:Ramond-Siegel_condition}
(1-\Prj_G) \nket{\PRamond} =0. 
\end{equation}

Because of the equations of motion 
\begin{equation}
\Prj_G \nket{\PNL_{(2+g)}} = \Prj_G \QB \nket{\PPRamond_{(1+g)}} 
\qquad \text{for}\quad g\geq0
\end{equation}
derived by varying the action \eqref{eq:Sgaugefixed-Ramond} with respect to $\nket{\PRamond_{\perp}{}_{(1-g)}}=(1-\Prj_G)\nket{\PRamond_{(1-g)}}$ the BRST transformations \eqref{eq:BRSTtrf_FP-aFP-NL} can be written in a single BRST transformation law:
\begin{equation} \label{eq:BRST-PRamond}
\delB \nket{\PPRamond} = \Prj_G \QB \nket{\PPRamond}.
\end{equation}

\subsection{Batalin-Vilkovisky (BV) action}

The iterative procedure discussed in the previous subsection is transparent and straightforward but is not applicable in an interacting case.
We need to use the BV formalism to fix the nonlinear gauge symmetry \eqref{eq:gaugetrfQB} of the interacting theory.\cite{Thorn:1986qj,Bochicchio:1986} 
Therefore, here, we first apply the BV formalism in the free R string case and show that it reproduces the results obtained in the previous subsection.

In the BV formalism, we require an extended action $S(\varphi,\varphi^*)$ to satisfy the BV master equation (or Zinn-Justin equation):
\begin{equation}
\sum_i
{\partial S\over\partial\varphi^i}{\partial S\over\partial\varphi_i^*} = 0,
\label{eq:BV-ME}
\end{equation}
where $\varphi_i^*$ denotes the antifield conjugate to the field $\varphi^i$.
If we have such an action, 
then the gauge fixing can be simply performed by setting the antifields 
$\varphi ^*_i$ equal to zero.
Note that we are adopting the BV formalism in the so-called gauge-fixed basis.\footnote{This is because the BV master action in this basis has the same form as the classical gauge-invariant action, as we will see shortly.}
The resultant gauge-fixed action is 
given by 
\begin{equation} \label{eq:BV-gaugefixe-action}
S_{\rm GF}(\varphi ) = S(\varphi,\varphi^*=0),
\end{equation}
which is invariant under the (true) BRST transformation 
\begin{equation}
\delta_\B \varphi^i = \left.{\partial S\over\partial\varphi^*_i}\right|_{\varphi^*=0}. 
\label{eq:BRST-trf-gene}
\end{equation} 
Indeed, it follows from the BV master equation that 
\begin{equation}
\delta_\B S_{\rm GF}(\varphi)
=\sum_i\left.{\partial S\over\partial\varphi^i}\right|_{\varphi^*{=}0}
\hspace{-0.5em}
\cdot \delta_\B \varphi^i
\hspace{0.5em}
=
\hspace{0.5em}
\left[ \sum_i{\partial S\over\partial\varphi^i}\,{\partial S\over\partial\varphi^*_i}\right]_{\varphi^*=0} =0.
\label{eq:BVmaster-eq}
\end{equation}

Let us apply the BV formalism to the free R string field theory.
In the previous subsection, it was shown that the gauge-fixed action has the same form as the classical gauge-invariant action without the ghost number constraint.
This suggests that the BV master action also takes the same form:
\begin{equation}
S = -\frac12 \fbra{\bPRamond} Y_0 \QB \fket{\PRamond}, \label{eq:master-action}
\end{equation}
which is invariant under the BRST transformation
\begin{equation} 
\delta_\B \nket{\PRamond} = \QB \nket{\PRamond}.
\end{equation}
We now show that this invariance implies that the action \eqref{eq:master-action} satisfies the BV master equation \eqref{eq:BVmaster-eq}.
Using the component form of the projected fields
\begin{eqnarray}
\nket{\PRamond}=
\dket{\phi} - \left(\gamma_0+c_0F\right) \dket{\psi},
\end{eqnarray}
we can calculate the inner product with the metric $Y_0$ as
\begin{align}
\fbra{\bPRamond_1} Y_0 \fket{\PRamond_2} 
&= \big( \dbra{\bar{\phi}_1} - \dbra{\bar{\psi}_1}\gamma_0 \big) c_0 \delta'(\gamma_0)
   \big( \dket{\phi_2} - \gamma_0 \dket{\psi_2}\big) 
\nn
&= \dbra{\bar{\phi}_1} c_0\delta(\gamma_0) \dket{\psi_2} 
 + \dbra{\bar{\psi}_1} c_0\delta(\gamma_0) \dket{\phi_2}
\nn
&= \nbraket{\bar{\phi}_1}{\psi_2} + \nbraket{\bar{\psi}_1}{\phi_2},
\label{eq:innerproductY_0}
\end{align}
where we have used Eq.~\eqref{eq:dket-ket}.
From Eq.~\eqref{eq:innerproductY_0}, we can see that the metric $Y_0$ is just the cross-diagonal nondegenerate metric between the two independent components $\ket{\phi}$ and $\ket{\psi}$ for the projected R field $\nket{\PRamond}$, just like $c_0$ in the bosonic string case.
Therefore, the general variation $\delta S$ can be written as
\begin{equation}
\delta S = - \fbra{\delta{\bar \PRamond}} Y_0 \QB \fket{\PRamond}
= - \fbra{\delta{\bar \PRamond}} Y_0 \fket{\delta_\B \PRamond} =
\nbraket{\delta{\bar \phi}}{\delta_\B\psi} + \nbraket{\delta{\bar \psi}}{\delta_\B\phi},
\end{equation}
with
\begin{equation}
\delta_\B \nket{\PRamond} =
\dket{\delta_\B\phi} - \left(\gamma_0+c_0F\right) \dket{\delta_\B\psi}.
\end{equation}
We thus have\footnote{The fields $\phi$ and $\psi$ in Eqs.~\eqref{eq:delB-ddS} actually represent the coefficient fields appearing in the expansion of the string states $\ket{\phi}$ and $\ket{\psi}$, respectively. Since $\nbra{\delta \bar \phi}= \bra{\delta \phi} \psi^0_0$ and $\psi^0_0$ is fermionic, these coefficient fields $\phi$ and $\psi$ have a nonvanishing inner product between those with opposite statistics, as it should be for the BV field and antifield. Although $\bar \phi = \phi^T C$ for the coefficient fields, we take $C=1$, which is possible for GSO-projected chiral fields.}
\begin{equation} \label{eq:delB-ddS}
{\partial S\over\partial{\psi}} = \delta_\B\phi, \qquad 
{\partial S\over\partial{\phi}} = \delta_\B\psi. 
\end{equation}
As a result, the BRST invariance of the master action implies that
\begin{equation}
0= \delta_\B S = {\partial S\over\partial\phi}\delta_\B\phi+{\partial S\over\partial\psi}\delta_\B\psi =
 2{\partial S\over\partial\phi}{\partial S\over\partial\psi},
\end{equation}
which is nothing but the BV master equation, 
if we identify $\phi$ and $\psi$ as the field and antifield, respectively.
In this identification, the BV gauge fixing defined by setting the antifield equal to zero gives the Ramond-Siegel gauge $\ket{\psi}=0$.
The BV gauge-fixed action \eqref{eq:BV-gaugefixe-action} and the BRST transformation
\begin{equation}
\delB \phi = \left. \frac{\dd S}{\dd \psi} \right|_{\psi=0} = \tQB \phi
\end{equation}
respectively coincide with the previous results \eqref{eq:Sgaugefixed-Ramond} and \eqref{eq:BRST-PRamond}.

\subsection{Propagator}

Before closing this section, we derive the propagator of the R string for the gauge condition \eqref{def:Ramond-Siegel_gauge}.
Let us consider the gauge-fixed free action \eqref{eq:Sgaugefixed-Ramond} with a source $J$:
\begin{align}
S[\PPRamond, J] &= -{1\over2}\fbra{\bPPRamond} Y_0 \QB \fket{\PPRamond} + \fbraket{\bPPRamond}{J}
\nn
&= - {1\over2}\nbra{{\bar \Ramond}} \Prj_Y^T \Prj_G^T Y_0 \QB \Prj_G \Prj_Y \ket{\Ramond} 
 + \nbra{{\bar \Ramond}} \Prj_Y^T \Prj_G^T \ket{J}.
\label{eq:S-R-J}
\end{align}
The propagator can be found by eliminating $\nket{\PPRamond}$ using the following equation of motion:
\begin{equation}
\Prj_Y^T \Prj_G^T \left( Y_0 \QB \nket{\PPRamond} - \ket{J} \right) =0.
\end{equation} 
A solution to this is given by
\begin{equation}
\nket{\PPRamond} = {b_0 X_0 \over L_0} \ket{J} 
 = b_0 {\delta(\beta_0) \over F_0} \ket{J}. 
\end{equation}
Indeed, using the relation \eqref{eq:PgdaggerY_0} and $F_0^2 = L_0$,
we can confirm this as follows:
\begin{align}
\Prj_G^T Y_0 \QB {b_0 X_0 \over L_0} \ket{J}
&= c_0 \delta(\gamma_0) \beta_0 \QB \frac{b_0 \delta(\beta_0)}{F_0} \ket{J}
\nn
&= c_0 \delta(\gamma_0) (-F_0) b_0 \delta(\beta_0) \frac{1}{F_0} \ket{J}
\nn
&= c_0 b_0 \delta(\gamma_0) \delta(\beta_0) \ket{J} 
 = \Prj_G^T \ket{J}.
\end{align}
Substituting this expression back into the action \eqref{eq:S-R-J}, we obtain
\begin{align}
S[\PPRamond(J), J] 
&= - \frac12 \nbra{{\bar J}} \frac{X_0 b_0}{L_0} Y_0 \QB \frac{b_0 X_0}{L_0} \ket{J}
 + \nbra{{\bar J}} \frac{X_0 b_0}{L_0} \ket{J}
\nn
&= - \frac12 \nbra{{\bar J}} {b_0X_0\over L_0} \ket{J}.
\end{align}
Thus, the Ramond propagator $\PiR$ is given by
\begin{equation} \label{def:R-propagator}
\PiR = {b_0X_0\over L_0} = b_0{\delta(\beta_0)\over F_0}.
\end{equation}

\section{Reduced form of NS field and BV master equation} 
\label{sec:NS_reducedform-BV}

We concentrate in this section on the NS sector action:
\begin{equation}
S = \half\int\YY\,\Phi * \QB \Phi + 
{1\over3} \int\YY\, \Phi * \Phi * \Phi. 
\end{equation}

\subsection{Resolving the kernel problem of $\YY$}

Owing to the kernel of the inverse picture-changing operator $\YY$,
if we expand the NS fields in powers of 
the ghost and superghost factors, $c(i),\ c(-i)$ and $\gamma(i),\ \gamma(-i)$, 
the terms containing $c(i),\ c(-i)$ and $\gamma^n(i),\ \gamma^n(-i)\ (n\geq2)$ 
can all be eliminated. 
In connection with this, we here define the following combinations of 
the ghost and superghost variables $c(\pm i)$ and $\gamma(\pm i)$ and 
note their mode expansions: 
\begin{eqnarray}
c_{\pm}(i) &\equiv&
\frac12 
\begin{pmatrix}
c(i) + c(-i)  \\
\ii^{-1}( c(i) - c(-i)) 
\end{pmatrix} =
\begin{pmatrix}
c_-  \\ 
c_0 \\
\end{pmatrix} + C_{\pm}(i) 
= \Exp{-T}
\begin{pmatrix}
c_-  \\ 
c_0 \\
\end{pmatrix}\Exp{T},
\nn
\gamma_{\pm}(i) &\equiv& 
\frac12 
\begin{pmatrix}
\gamma(i) + \gamma(-i)  \\
\ii^{-1}( \gamma(i) - \gamma(-i)) 
\end{pmatrix} =
\gamma_{\pm\half}+ \Gamma_{\pm}(i) = \Exp{-T}
\begin{pmatrix}
\gamma_{+\half} \\ 
\gamma_{-\half} \\
\end{pmatrix} \Exp{T},
\end{eqnarray}
where $c_- = c_1 - c_{-1}$, and
\begin{eqnarray}
T &=& b_0 C_- + \half b_- C_+ + \beta_{-\half} \Gamma_+ + 
\beta_{\half} \Gamma_-, 
\qquad 
b_{-} = \frac12 (b_{-1} - b_1),
\label{def:translationT}
\\
C_{+}(i) &=& \sum_{k=1}^\infty(-1)^k (c_{2k+1}-c_{-(2k+1)}), \quad 
C_{-}(i) = \sum_{k=1}^\infty(-1)^k (c_{2k}+c_{-2k}), \nn
\Gamma_{+}(i) &=&  \sum_{k=1}^\infty(-1)^k (\gamma_{2k+\half}+\gamma_{-2k+\half}), \quad 
\Gamma_{-}(i) = \sum_{k=1}^\infty(-1)^k (\gamma_{2k-\half}+\gamma_{-2k-\half}).
\end{eqnarray}
Then, in front of the measure $\YY$, 
the general NS string field $\Phi$ can always 
be rewritten into the following projected form  
(noting that $\gamma^2_{+}(i)+\gamma^2_{-}(i)=\gamma(i)\gamma(-i)$):\footnote{These states are therefore the tensor product of the superconformal vacuum $\ket{0}$ in the $(\beta_{\pm\frac12},\gamma_{\pm\frac12};b_{0,\pm1},c_{0,\pm1})$ sector and the states $\ket{\phi_0}, \ket{\psi_0},\cdots$ in the other mode sector. Therefore, if we follow the notation in \S \ref{sec:gaugefix-R}, these states should be written as $\dket{\phi_0}, \dket{\psi_0}, \cdots$, but here we omit the distinction between $\ket{~~}$ and $\dket{~~}$ for simplicity. }
\begin{align}
\Prj_0 \Phi&\equiv\widetilde\Phi
\nn
&\equiv
\ket{\phi_0}+\gamma_+(i)\ket{\phi_+}+\gamma_{-}(i)\ket{\phi_-}
+\half(\gamma^2_{+}(i)+\gamma^2_{-}(i))\ket{\phi_{+-}} 
\nn
&\quad + c_+
\Bigl(\ket{\psi_0}+\gamma_+(i)\ket{\psi_+}+\gamma_{-}(i)\ket{\psi_-}
+\half(\gamma^2_{+}(i)+\gamma^2_{-}(i))\ket{\psi_{+-}} \Bigr),
\label{eq:P0Phi}
\end{align}
where $c_{+} \equiv c_1+c_{-1}$ 
and all the component fields are annihilated by 
$b_0, b_{\pm1}, \beta_{\pm\half}$:
\begin{equation}
\bigl( b_0,\ b_{\pm1},\ \beta_{\pm\half} \bigr) 
\bigl( \ket{\phi_0},\ \ket{\phi_{\pm}},\ \ket{\phi_{+-}},\ \ket{\psi_0}, 
\ket{\psi_{\pm}}, \ket{\psi_{+-}} \bigr) =0. 
\end{equation} 
Since $\Exp{T}\gamma_{\pm}(i)\Exp{-T}= \gamma_{\pm\half}$,  
the expansion in Eq.~\eqref{eq:P0Phi} is in fact the mode expansion 
in powers of $\gamma_{\pm\half}$ if transformed by the translation 
operator $\Exp{T}$:
\begin{eqnarray}
\Exp{T} \Prj_0 \Phi &=& 
\ket{\phi_0}+\gamma_{+\half}\ket{\phi_+}+\gamma_{-\half}\ket{\phi_-}
+\half(\gamma^2_{+\half}+\gamma^2_{-\half})\ket{\phi_{+-}} \nn
&&{}+c_+\Bigl( \ket{\psi_0}+\gamma_{+\half}\ket{\psi_+}+\gamma_{-\half}\ket{\psi_-}
+\half(\gamma^2_{+\half}+\gamma^2_{-\half})\ket{\psi_{+-}}\Bigr).
\end{eqnarray}
The explicit expression of the projection operator $\Exp{T}\Prj_0\Exp{-T}$ 
can be given as 
\begin{eqnarray}
\Exp{T}\Prj_0\Exp{-T}
&=& b_- c_- 
 \bigg[ 
  \delta^2(\beta)\delta^2(\gamma) 
  - \gamma_{\half} \delta^2(\beta)\delta^2(\gamma) \beta_{-\half}
  - \gamma_{-\half} \delta^2(\beta)\delta^2(\gamma) \beta_{\half}
\nn
&&\qquad\qquad\qquad\quad
  + \half\left(\gamma_{\half}^2 + \gamma_{-\half}^2 \right) \delta^2(\beta)\delta^2(\gamma) \half \left(\gamma_{\half}^2 + \gamma_{-\half}^2 \right)
 \bigg],
\\
&&\delta^2(\beta) = \delta(\beta_{\half}) \delta(\beta_{-\half}), 
\quad 
\delta^2(\gamma) = \delta(\gamma_{\half}) \delta(\gamma_{-\half}).
\end{eqnarray}

In front of the $\YY$, we can always reduce the field into the 
projected form \eqref{eq:P0Phi}. We adopt this reduction as our 
convention for the NS field,
 then the kernel problem of $\YY$ is resolved. 
The operator $\YY$ in fact gives a nondegenerate 
and cross-diagonal metric in this 
projected subspace. 
Indeed, noting that
\begin{equation}
{} \Exp{T} \YY \Exp{-T} \equiv 
(\YY)_0 = \frac14 c_- c_0 
\left(
\delta''(\gamma_{+\half})\delta(\gamma_{-\half}) + 
\delta(\gamma_{+\half})\delta''(\gamma_{-\half}) 
\right),
\end{equation}
and 
$\delta''(\gamma)=\bigl[ [\delta(\gamma),\,\beta],\,\beta] = \beta^2\delta(\gamma)-2\beta\delta(\gamma)\beta+\delta(\gamma)\beta^2$
for $\gamma=\gamma_{\pm\half}$ and $\beta=\beta_{\mp\half}$ satisfying $[\gamma,\,\beta]=1$, 
we find that the inner product of the two NS fields $\Phi^{(1)}$ and $\Phi^{(2)}$
written in the form (\ref{eq:P0Phi}) is given by
\begin{eqnarray} \label{eq:cross-diag_NS}
&&\bra{\Phi^{(1)}}\YY \ket{\Phi^{(2)}} =
\big\langle{\widetilde\Phi^{(1)}}\big| \YY \big|{\widetilde\Phi^{(2)}}\big\rangle= 
\big\langle\tilde\Phi^{(1)}\big| \Exp{-T}(\YY)_0 \Exp{T} \big|\tilde\Phi^{(2)}\big\rangle\nn
&& \hspace{1.5em}=
\big\langle{\phi^{(1)}_0}\big|{\psi^{(2)}_{+-}}\big\rangle 
+\big\langle{\phi^{(1)}_+}\big|{\psi^{(2)}_{-}}\big\rangle 
+\big\langle{\phi^{(1)}_-}\big|{\psi^{(2)}_{+}}\big\rangle 
+\big\langle{\phi^{(1)}_{+-}}\big|{\psi^{(2)}_0}\big\rangle 
+ ( \phi\leftrightarrow \psi).
\end{eqnarray}
This cross-diagonal form of the inner product is very important for the 
BV master equation below.

\subsection{BV action satisfying the master equation}

We claim that the action satisfying the BV master equation takes the same form as the classical gauge-invariant action with the ghost number constraint relaxed. 

The gauge invariance of the original classical action was realized by the 
derivation property and partial integrability of the BRST operator $\QB$ 
on the star product as well as the associativity of the 
star product. 
Then the present action with ghost number constraint on $\Phi$ relaxed can also be seen as invariant under the following BRST transformation $\delta_\B$:
\begin{equation}
\preBRST \Phi=  \QB \Phi+ \Phi * \Phi.
\end{equation} 
The action $S$ is constructed such that it takes a particular form for any variation $\delta\Phi$:
\begin{equation}
\delta S = \int \YY \delta\Phi* \delB\Phi= \bra{\delta\Phi}\YY \ket{\delta_\B \Phi}.
\label{eq:ActionForm}
\end{equation}
Note that the presence of $\YY$ does not injure the partial integrability 
and derivation property of the BRST operator $\QB$ because $[\QB, \YY]=0$ 
nor the associativity of the star product since $\YY$ is placed at 
the midpoint that is common to all the participating strings. 
Therefore, the action is invariant under the BRST transformation
\begin{equation}
\delta_\B S = \int \YY \delta_\B \Phi* \delta_\B\Phi= 0.
\label{eq:BRSTinvariance}
\end{equation}

However, it is not quite trivial that this BRST invariance of the 
system guarantees the BV master equation of the action $S$ because of the 
presence of the $\YY$ factor. 
It has the kernel so that we projected the NS field $\Phi$ into the 
reduced form 
$\Prj_0\Phi=\widetilde\Phi$, as given in Eq.~\eqref{eq:P0Phi}. 
But the above BRST 
transformation $\delta_\B\Phi$ does {\em not} 
give a closed transformation in such a space of reduced form 
string fields $\widetilde\Phi$. 
Fortunately, however, the $\YY$ factor in the action remains present in front 
even after the BRST transformation, so that it is 
automatically projected into the reduced form. That is, we can define 
the BRST transformation in the reduced space as 
\begin{equation}
\delta_\B \widetilde\Phi 
= \Prj_0 
(\QB \widetilde\Phi+ \widetilde\Phi*\widetilde\Phi).
\end{equation}
Here, we have written all $\Phi$ by $\widetilde\Phi$ since we regard 
the reduced field $\widetilde\Phi$ as our basic variable, and actually only 
those components appear in the action because of the presence of the 
$\YY$ factor.\footnote{This, in fact, holds even in the absence of the $\YY$ factor for the interaction term, since  the ghost $c(\pm i)$ and superghost $\gamma(\pm i)$ at the midpoint vanish on Witten's three-string vertex even without the $\YY$ factor, Eq.~\eqref{eq:vprop}.}
Equation \eqref{eq:ActionForm} can be rewritten in the form
\begin{equation}
\delta S = \nbra{\delta\widetilde\Phi} \YY \nket{\delta_\B \widetilde\Phi\,}
= \big\langle{ \Exp{T}\delta\widetilde\Phi }\big| (\YY)_0 \big|{\Exp{T} \delta_\B \widetilde\Phi}\big\rangle.
\end{equation}
However, we know that the metric $\YY$ is cross-diagonal in the reduced 
NS string field space, so that we have 
\begin{equation}
\delta S = 
\big\langle{\delta\phi_0}\big|{\delta_\B\psi_{+-}}\big\rangle 
+\big\langle{\delta\phi_+}\big|{\delta_\B\psi_{-}}\big\rangle 
+\big\langle{\delta\phi_-}\big|{\delta_\B\psi_{+}}\big\rangle 
+\big\langle{\delta\phi_{+-}}\big|{\delta_\B\psi_0}\big\rangle 
+ ( \phi\leftrightarrow \psi);
\label{eq:metricStr}
\end{equation} 
hence,
\begin{equation}
\delta_\B\psi_i= {\partial S\over\partial\phi_{I(i)}},
\quad 
\delta_\B\phi_i= {\partial S\over\partial\psi_{I(i)}},
\qquad 
\hbox{with} \quad 
i= 
\begin{pmatrix}
+- \\ \mp \\ 0 
\end{pmatrix}
\ \leftrightarrow \ 
I(i)=
\begin{pmatrix}
0 \\ \pm\\ {+-} 
\end{pmatrix}.
\end{equation}
With these relations, the BRST invariance of the action implies 
that the action $S$ satisfies the BV master equation; indeed, we have 
\begin{equation}
0=\delta_\B S = \!\!\sum_{i= 0, \pm, +-}\!\!\left(
{\partial S\over\partial\phi_i}\,\delta_\B\phi_i+
{\partial S\over\partial\psi_i}\,\delta_\B\psi_i\right)
= \sum_i\left(
{\partial S\over\partial\phi_i}\,{\partial S\over\partial\psi_{I(i)}}+
{\partial S\over\partial\psi_i}\,{\partial S\over\partial\phi_{I(i)}}\right),
\end{equation}  
so that
\begin{equation}
\sum_i
{\partial S\over\partial\phi_i}\,{\partial S\over\partial\psi_{I(i)}} =0.
\end{equation}

\section{Gauge fixing of NS sector by component fields}
\label{sec:gf-component-NS}

Now that we have shown that our action $S$ satisfies the BV master 
equation, we can discuss how to fix the gauge explicitly in 
component fields.

The gauge can generally be fixed by setting the antifields equal to zero.
However, the choice of the gauge-fixing conditions is of course not unique. 
Accordingly, the choice of the set of the antifields is also not 
unique, and there is actually the freedom of doing graded canonical 
transformation of field and antifield variables. 
However, we should note that the latter freedom of 
graded canonical transformation is much wider\footnote{Indeed, even exchanging the fields and antifields is contained as a special graded canonical transformation, as is well known.} 
and that {\em not} all the sets of antifields can be used as gauge-fixing conditions. 
To find suitable sets of antifields to be set equal to zero as gauge fixing, we have to confirm that they can actually be set equal to zero by the BRST transformation 
(which contains the original gauge transformation for the original gauge fields). 

To understand the situation better, let us recall the case of bosonic 
SFT as the simplest example of gauge fixing.
The bosonic string 
field is expanded in the FP ghost zero mode $c_0$ as $\Phi= \phi+c_0 \psi$ 
and the action $S$ satisfies the BV master equation
\begin{equation}
{\partial S\over\partial\phi}\,{\partial S\over\partial\psi} =0.
\end{equation}
Since this BV master equation is totally symmetric between $\phi$ and $\psi$, 
we cannot determine from this equation alone which variable $\phi$ or $\psi$ 
should be taken as the antifield set to be zero. 
To determine it properly, we need to 
have a closer look into the (free part of the) BRST transformation 
$\delta_\B\Phi=\QB\Phi=(c_0 L_0 + b_0 M + \tQB) \Phi$, which takes the forms of the component fields $\phi$ and $\psi$:
\begin{eqnarray}
\delta_\B\phi&=& \tQB \phi+ M \psi, \\
\delta_\B\psi&=& -L_0 \phi+ \tQB \psi.
\end{eqnarray}
From this expression, we understand why we can take $\psi$ but not $\phi$ 
as the antifield, since $\psi$ can be set equal to zero using the 
$-L_0\phi$ part in $\delta_\B\psi$. Note that the Klein-Gordon Virasoro operator 
$L_0$ is regarded as invertible in this discussion of general off-shell 
fields.%
\footnote{Even the on-shell $L_0=0$ component of $\psi$ can be gauged away by using the `dipole-ghost' component in $\phi$, satisfying $L_0\phi=-\psi\not=0$ and $L_0^2\phi=0$.} 
This is the moral of the game that we fully use below. 
However, the operators $\tQB$ and $M$ appearing on the RHS of $\delta_\B\phi$ 
are non-invertible; thus, $\phi$ cannot totally be gauged away. 
In addition, we know that the physical gauge field is contained in $\phi$, which also supports the observation that $\phi$ cannot be eliminated.

Now we go back to our discussion of the present problem of the NS string. 

We first show that the $b_+\widetilde\Phi=0$ gauge adopted by PTY is invalid: 
\begin{equation}
b_+\widetilde\Phi=0 \quad \leftrightarrow \quad 
\psi_{0}=\psi_{+}=\psi_{-}=\psi_{+-}=0.
\end{equation}
The reason why this gauge is not good is that not only is its perturbation theory very singular, but the gauge itself cannot be taken. 
Indeed, the $\psi_{0}$ component contains, for instance, the 
original (ghost-number zero) physical gauge field, which clearly cannot be gauged away. 
As was shown by Urosevic and Zubarev \cite{Urosevic:1990as} explicitly, $\ket{\psi_0}$ contains 
\begin{eqnarray}
\ket{\psi_0} &=&  
\half A_\mu(k)\alpha_{-1}^\mu\Exp{ikx}\ket{0},
\end{eqnarray}
which is identified as the massless gauge field.
Indeed the BRST transformation 
$\delta_\B\widetilde \Phi=\Prj_0\QB \widetilde \Phi$ for the component  
$\ket{\psi_0}$ is given by
\begin{equation}
\delta_\B\ket{\psi_0}= -L_+ \ket{\phi_0} + \cdots, 
\end{equation}
but only the part $\half p_\mu\alpha^\mu_{-1}$ in $L_{+}$ with 
$\ket{\phi_0}= \lambda(k)\Exp{ikx}\ket{0}$ can contribute to the transformation 
and takes the form of the conventional gauge transformation
\begin{equation}
\delta_\B A_\mu(k) = k_\mu\lambda(k).
\end{equation}
Namely, the transverse part of the gauge field $A_\mu$ at least cannot be 
gauged away. This also demonstrates that the impossibility of eliminating $\ket{\psi_0}$ is essentially connected to the non-invertibility of 
the operator $L_+$. 
If it were invertible, the $-L_+ \ket{\phi_0}$ part in $\delta_\B\ket{\psi_0}$ could have eliminated everything in $\ket{\psi_0}$. 
Therefore this explains the reason why the impossibility of the $b_+=0$ gauge is related to the singular perturbation theory using the propagator $b_+/L_+$. 

Now, we consider the BRST transformation 
$\delta_\B\widetilde\Phi$ for all the eight 
component fields 
$\ket{\phi_{0}}, \ket{\phi_{\pm}}, \ket{\phi_{+-}}$ and 
$\ket{\psi_{0}}, \ket{\psi_{\pm}}, \ket{\psi_{+-}}$, the explicit form of 
which is given in Appendix \ref{app:BRSTop-trf}. 
Since it is simpler if we eliminate higher-power terms in 
$\gamma_{\pm\half}$, we first try to eliminate $\ket{\psi_{+-}}$.  
We note that the RHS of $\delta_\B \ket{\psi_{+-}}$ contains 
$(\Gpsim)/2$ so, if $\ket{\psi_{\pm}}$ contain the component of the 
form $G_{\pm\half}\ket{\psi}$ with a common $\ket{\psi}$, then 
it yields 
\begin{equation}
\half\left(\Gpsim \right) = 
\half \bigl\{G_{-\half}, \, G_{+\half}\bigr\}\ket{\psi}= L_0\ket{\psi}.
\end{equation}
Since $L_0$ is invertible as emphasized above, this freedom $L_0\ket{\psi}$ 
can totally eliminate the $\ket{\psi_{+-}}$ component. In the same way, 
we can eliminate $\ket{\phi_{+-}}$ using the part 
\begin{equation}
\half \left( G_{-\half}\ket{\phi_{+}}+G_{\half}\ket{\phi_{-}} \right)
\quad \hbox{or}\quad 
-2\ket{\psi_0}
\end{equation}
contained in $\delta_\B \ket{\phi_{+-}}$. 
Thus, we have seen that we can take the gauge
\begin{equation} \label{eq:gaugecond-NS-1}
\ket{\psi_{+-}}=\ket{\phi_{+-}}=0\ .
\end{equation}
Under these gauge conditions, 
$\ket{\psi_{+-}}$ and $\ket{\phi_{+-}}$ are now identified as antifields and 
the components $\ket{\phi_{0}}$ and $\ket{\psi_{0}}$ are the fields 
conjugate to them, respectively, as seen 
from the metric structure (\ref{eq:metricStr}).

Next, looking at the transformation law $\delta_\B \ket{\psi_{\pm}}$, we find 
the part $\pm\half\ket{\phi_{\mp}}$, so that we are tempted to choose the 
gauge setting $\ket{\psi_{\pm}}=0$. 
However, if we do so, all the terms on 
the RHS of $\delta_\B \ket{\psi_{+-}}$ vanish. 
This implies by Eq.~(\ref{eq:metricStr}) that the field 
component $\ket{\phi_{0}}$ does not appear at all in the kinetic term in the action. 
Thus the kinetic term is singular so 
that it is not allowed. This observation is consistent with the statement 
that we should keep the component $\Gpsim$ nonzero so that we can 
use it to eliminate the $\ket{\psi_{+-}}$ component. 
The lesson here is that the RHS of the BRST transformation of the 
{\em antifield}, giving an equation of motion of the gauge-fixed action,  
should contain {\em field} components not eliminated by gauge fixing.

To find suitable gauge-fixing conditions for the sector of 
$(\ket{\phi_{\pm}}, \ket{\psi_{\pm}})$, let us introduce 
the following decomposition of the components 
$\ket{\phi_{\pm}}$ into $\ket{\phi}$ and $\ket{\phi^*}$: 
\begin{equation} \label{eq:phipm-decomp}
\ket{\phi_{\pm}} = G_{\pm\half}\ket{\phi} + G_{\mp\half}\ket{\phi^*}.
\end{equation}
Conversely, 
these $\ket{\phi}$ and $\ket{\phi^*}$ can be expressed 
in terms of $\ket{\phi_{\pm}}$ as follows.
Multiplying $G_{\pm\frac12}$ to \eqref{eq:phipm-decomp} and using $\big(G_\frac12\big)^2 + \big(G_{-\half}\big)^2 = 2L_+$ and $\{G_{\half},G_{-\half}\}= 2L_0$, we obtain
\begin{eqnarray}
\ket{G_{\mp}\phi_{\pm}} &=& L_0 \ket{\phi}+ L_+ \ket{\phi^*}, \nn
\ket{G_{\pm}\phi_{\pm}} &=& L_+ \ket{\phi}+ L_0 \ket{\phi^*},
\label{eq:phistar-phipm}
\end{eqnarray}
where we introduced the notation
\begin{equation} \label{def:Gphi}
\fket{G_{\pm}\phi_{\pm}}\equiv\Gphip,\qquad 
\fket{G_{\mp}\phi_{\pm}}\equiv\Gphim.
\end{equation}
Since $L_0$ is invertible, Eq.~\eqref{eq:phistar-phipm} can be solved as
\begin{eqnarray}
\Bigl[ 1- \Bigl(L_+\frac{1}{L_0} \Bigr)^2\Bigr] 2L_0 \ket{\phi} 
&=& 
\ket{G_{\mp}\phi_{\pm}} - L_+\frac{1}{L_0} \ket{G_{\pm}\phi_{\pm}}, \nn
\Bigl[ 1- \Bigl(L_+\frac{1}{L_0} \Bigr)^2\Bigr] 2L_0 \ket{\phi^*} 
&=& 
\ket{G_{\pm}\phi_{\pm}} - L_+\frac{1}{L_0} \ket{G_{\mp}\phi_{\pm}}.
\end{eqnarray}
The symplectic metric in this sector
\begin{equation}
\big\langle{\phi_+}\big|{\psi_{-}}\big\rangle 
+\big\langle{\phi_-}\big|{\psi_{+}}\big\rangle 
=
\big\langle{\phi}\big|\,G_{\pm}\psi_{\pm}\big\rangle 
+\big\langle{\phi^*}\big|\, G_{\mp}\psi_{\pm}\big\rangle,
\end{equation}
with the same notation for $\ket{\psi_{\pm}}$ as Eq.~\eqref{def:Gphi}, 
implies that if we take the component $\ket{G_{\mp}\psi_{\pm}}$
as a {\em field}, then $\ket{\phi^*}$ is the conjugate antifield; indeed, 
\begin{equation}
\delta_\B\ket{\phi_{\pm}}= G_{\mp\half}\ket{\phi_0}+\cdots
\end{equation} 
implies that we can eliminate $\ket{\phi^*}$ in $\ket{\phi_{\pm}}$ by the freedom of $\ket{\phi_0}$. 
However, this BRST transformation also shows that 
another freedom $\ket{\phi}$ in $\ket{\phi_{\pm}}$ can no longer 
be eliminated so that $\ket{\phi}$ should be a {\em field} 
and $\fket{G_{\pm}\psi_{\pm}}$ must be an antifield. 
The BRST transformation of $\fket{G_{\pm}\psi_{\pm}}$ is given by
\begin{equation}
\delta_\B \fket{G_{\pm}\psi_{\pm}} = 2L_0\ket{\psi_0}+\cdots, 
\end{equation}
so that it can actually be eliminated. 

We are thus led to the following gauge-fixing conditions in 
the $(\ket{\phi_{\pm}}, \ket{\psi_{\pm}})$ sector:
\begin{equation} \label{eq:gaugecond-NS-2}
\fket{G_{\pm}\psi_{\pm}}=0, \quad \ket{\phi^*}\propto 
\ket{G_{\pm}\phi_{\pm}} - L_+\frac{1}{L_0} \ket{G_{\mp}\phi_{\pm}}=0;
\end{equation}
the remaining {\em field} degrees of freedom are 
\begin{equation}
\ket{\phi}\qquad \hbox{and}\qquad   
\ket{G_{\mp}\psi_{\pm}}\ .
\end{equation}

We have thus identified gauge-fixing conditions as well as the fields and antifields.
We can now read out the BRST transformation \eqref{eq:BRST-trf-gene} of the fields after gauge fixing under which the gauge-fixed action is invariant.

So far so good. 
However,
we have not yet succeeded 
in finding the propagator in a closed form in this gauge. This is 
because the gauge-fixed action takes a rather complicated form 
in terms of the component fields. 
Therefore, in the next section, we try another approach to gauge fixing, 
directly working with total string fields, which gives gauge-fixing conditions that are very close to the above conditions in this section. 

\section{Gauge fixing of NS sector by nonlocal projection} 
\label{sec:NSgf_nonlocal-gc}

In this section, we consider another gauge-fixing approach that is not as explicit as the that discussed in the previous section but more suitable for computing the propagator.
We only investigate the free theory part which is sufficient for the purpose to study what conditions can be taken as the gauge choice.

\subsection{Iterative gauge fixing}

If we concentrate only on the free theory part, the conventional gauge-fixing procedure can be applied.
The free classical action of the NS sector is 
\begin{align}
S_0 &= \frac{1}{2}\langle\Phi_{(1)}|Y\bar{Y}Q_\B|\Phi_{(1)}\rangle
= \frac{1}{2}\langle\hat{\Phi}_{(1)}|Y\bar{Y}Q_\B|\hat{\Phi}_{(1)}\rangle,\label{eq:caction}
\end{align}
where $|\Phi_{(1)}\rangle$ is the classical string field restricted onto the ghost number one. 
Here, in this section, we project out the kernel degrees of freedom of $Y\bar{Y}$ as the hatted field $|\hat{\Phi}_{(1)}\rangle=\mathcal{P}_{Y\bar{Y}}|\Phi_{(1)}\rangle$ using the BRST-invariant projection operator
\begin{align}
\mathcal{P}_{Y\bar{Y}} = X_{-\frac{1}{2}}X_{\frac{1}{2}}Y\bar{Y},\qquad
[Q_\B, \mathcal{P}_{Y\bar{Y}}] = 0,
\label{eq:yyprojection}
\end{align}
where
\begin{align}
X_{\pm\frac{1}{2}} &= [Q_\B, \Theta(\beta_{\pm\frac{1}{2}})]
 = \delta(\beta_{\pm\frac{1}{2}})G_{\pm\frac{1}{2}}-b_{\pm1}\delta'(\beta_{\pm\frac{1}{2}}).
\end{align}
To compute the propagator, this projection operator is  more convenient than the $\mathcal{P}_0$ introduced in \S \ref{sec:NS_reducedform-BV}, although its component form is very complicated. 
Because the operator $X_{-\frac{1}{2}}X_{\frac{1}{2}}$ commutes with both
$L_0$ and $b_0$, i.e.,
\begin{equation}
\label{eq:xxrelation}
[L_0, \Xm\Xp] = 0, \qquad [b_0, \Xm\Xp] = 0,
\end{equation}
this is suitable for an analog of the conventional Siegel gauge, as we will see shortly.

The action (\ref{eq:caction}) is invariant under the gauge transformation
\begin{equation}
\delta|\hat{\Phi}_{(1)}\rangle = Q_\B |\hat{\Lambda}_{(0)}\rangle,\label{eq:gaugetf}
\end{equation}
so we must fix it by choosing an appropriate gauge condition.
For this purpose, we generally define a decomposition of the hatted NS field $\nket{\hat A} = \Prj_{\YY} \ket{A}$ by
\begin{align}
 |\hat{A}\rangle &= \mathcal{P}_{\rm NS}|\hat{A}\rangle + \mathcal{P}_{\rm NS}^\perp|\hat{A}\rangle \nn
&\equiv |\hat{A}_{\parallel}\rangle + |\hat{A}_{\perp}\rangle,
\label{eq:decompA}
\end{align}
where $\mathcal{P}_{\rm NS}$ and $\mathcal{P}_{\rm NS}^\perp$ are, respectively, the Siegel-gauge projection operators $\Prj_b$ and $(1-\Prj_b)$ restricted in the hatted subspace:
\begin{align}
\Prj_b &\equiv \frac{b_0}{L_0} \QB,
\label{def:Pb}
\\
\mathcal{P}_{\rm NS} 
&= \mathcal{P}_{Y\bar{Y}} \Prj_b \mathcal{P}_{Y\bar{Y}},
\qquad
 \mathcal{P}_{\rm NS}^\perp 
= \mathcal{P}_{Y\bar{Y}}(1-\mathcal{P}_b)\mathcal{P}_{Y\bar{Y}} = \mathcal{P}_{\rm NS}^\dag.
\end{align}
These operators satisfy
\begin{align}
&\mathcal{P}_{Y\bar{Y}} = \mathcal{P}_{\rm NS} + \mathcal{P}_{\rm NS}^\perp,\\
&(\mathcal{P}_{\rm NS})^2 = \mathcal{P}_{\rm NS},\qquad (\mathcal{P}_{\rm NS}^{\perp})^2 = \mathcal{P}_{\rm NS}^\perp,\\
&\mathcal{P}_{\rm NS}\mathcal{P}_{\rm NS}^\perp = \mathcal{P}_{\rm NS}^\perp\mathcal{P}_{\rm NS} = 0,\\
&\mathcal{P}_{Y\bar{Y}}^\dag Y\bar{Y}\mathcal{P}^\perp_{\rm NS} =
\mathcal{P}_{\rm NS}^\dag Y\bar{Y}\mathcal{P}_{Y\bar{Y}}.\label{eq:cdrelation}
\end{align}
This decomposition \eqref{eq:decompA} using the nonlocal projection operator $\Prj_b$ in \eqref{def:Pb} splits the hatted string field into halves because
the ranks of the projection operators $\mathcal{P}_{\rm NS}$ and $\mathcal{P}_{\rm NS}^\dag$ must be the same.
Using the last relation (\ref{eq:cdrelation}), which follows from the commutativity (\ref{eq:xxrelation}), one can show that this has a nondegenerate and cross-diagonal inner product:
\begin{equation}
 \langle \hat{A}_1|Y\bar{Y}|\hat{A}_2\rangle =
\langle \hat{A}_{1\perp}|Y\bar{Y}|\hat{A}_{2\parallel}\rangle
+ \langle \hat{A}_{1\parallel}|Y\bar{Y}|\hat{A}_{2\perp}\rangle.
\end{equation}
Although these facts are useful to identify the gauge condition and carry out iterative gauge fixing as we will see below, it should be noted that 
this decomposition cannot exactly be identified as the field and antifield decompositions as was carried out in the previous section in Eq.~\eqref{eq:cross-diag_NS}.
In this regard, the detailed analysis is given in Appendix \ref{app:comp-nonlocal}
in the R string case, which can be explicitly studied using component field expansion in ghost zero modes.

Note that the gauge transformation (\ref{eq:gaugetf}) splits in this decomposition into 
\begin{align}
 \delta |\hat{\Phi}_\parallel{}_{(1)}\rangle &= 0,\\
 \delta |\hat{\Phi}_\perp{}_{(1)}\rangle &= Q_\B|\hat{\Lambda}_\parallel{}_{(0)}\rangle,\label{eq:delPhiperp}
\end{align}
since $\Prj_{\rm NS} \QB =0$ and $\Prj_{\rm NS}^\perp \QB = \QB \Prj_{\rm NS}$.
Because of Eq.~\eqref{eq:delPhiperp}, we can gauge away the $|\hat{\Phi}_\perp{}_{(1)}\rangle$ part, that is, we can choose the gauge condition
\begin{equation}
 |\hat{\Phi}_\perp{}_{(1)}\rangle = \mathcal{P}_{Y\bar{Y}}(1-\mathcal{P}_b)|\hat{\Phi}_{(1)}\rangle = 0,
\label{eq:gfcondition}
\end{equation}
which must be equivalent to setting the antifield equal to zero.

The BRST transformation is defined by replacing the gauge parameter $ |\hat{\Lambda}_{(0)}\rangle$
by the ghost string field $|\hat{\Phi}_{(0)}\rangle$:
\begin{equation}
\delta_\B|\hat{\Phi}_{(1)}\rangle = Q_\B |\hat{\Phi}_{(0)}\rangle.
\end{equation}
To construct the gauge-fixed action, it is convenient to also introduce the antighost field 
$ |\hat{\Phi}_{(2)}\rangle$ and the NL field $ |\hat{B}_{(2)}\rangle$
with the BRST transformation
\begin{equation}
\delta_\B|\hat{\Phi}_{(2)}\rangle = |\hat{B}_{(2)}\rangle,\qquad
\delta_\B|\hat{B}_{(2)}\rangle = 0.
\end{equation}
The gauge-fixing and FP ghost action can be given by using these fields as
\begin{align}
S_1 &= -\delta_\B\left(\langle\hat{\Phi}_{(2)}|Y\bar{Y}(1-\mathcal{P}_b)|\hat{\Phi}_{(1)}\rangle\right)
\nn
&= -\langle \hat{B}_{(2)}|Y\bar{Y}(1-\mathcal{P}_b)|\hat{\Phi}_{(1)}\rangle
+\langle\hat{\Phi}_{(2)}|\mathcal{P}_b^\dag Y\bar{Y}Q_\B|\hat{\Phi}_{(0)}\rangle.
\end{align}
Here, we have used the relation (\ref{eq:cdrelation}).
We can eliminate the NL field and rewrite the total action in the first step in the form
\begin{equation}
S_0+S_1 = \frac{1}{2}\langle\hat{\Phi}_\parallel{}_{(1)}|Y\bar{Y}Q_\B|\hat{\Phi}_\parallel{}_{(1)}\rangle
+ \langle\hat{\Phi}_\parallel{}_{(2)}|Y\bar{Y}Q_\B|\hat{\Phi}_{(0)}\rangle.\label{eq:gfaction1}
\end{equation}
Here, it must be again noted that $|\hat{\Phi}_\parallel{}_{(1)}\rangle$ is not purely the field component but also contains the antifield component, although we do not need to consider it to compute the propagator. 
However, we must set its antifield component equal to zero to obtain the BRST transformation
of the (gauge-fixed) field because $\delta_\B|\hat\Phi_\parallel{}_{(1)}\rangle = 0$. (See Appendix \ref{app:comp-nonlocal} for detailed analysis in the R string case.)

The first step action (\ref{eq:gfaction1}) still has a new gauge invariance under the transformation
\begin{equation}
\delta|\hat{\Phi}_{(0)}\rangle = Q_\B |\hat{\Lambda}_{(-1)}\rangle,
\end{equation}
which can be similarly fixed by introducing a set of (ghost for) ghost, NL, and antighost fields,
$(|\hat{\Phi}_{(-1)}\rangle, |\hat{B}_{(3)}\rangle, |\hat{\Phi}_{(3)}\rangle)$.  We can repeat the same procedure
as the first step.
We must infinitely repeat the procedure; however, we can easily see that the final form of the gauge-fixed action is
\begin{equation}
S_{GF} = \frac{1}{2}\langle\hat{\Phi}_\parallel|Y\bar{Y}Q_\B|\hat{\Phi}_\parallel\rangle,\label{eq:totalgfaction}
\end{equation}
where $|\hat{\Phi}_\parallel\rangle$ defined by
\begin{equation}
|\hat{\Phi}_\parallel\rangle = \sum_{g=-\infty}^{\infty}|\hat{\Phi}_\parallel{}_{(g)}\rangle,
\end{equation}
is the gauge-fixed string field with the ghost number constraint relaxed.%
\footnote{Strictly speaking, we must set its antifield component equal to zero, as already mentioned.}

\subsection{Propagator}

We can also derive the propagator for the gauge condition (\ref{eq:gfcondition}). 
Let us start by adding a source term to the free gauge-fixed action (\ref{eq:totalgfaction}):
\begin{align}
S[\Phi, J] &= \frac{1}{2}\langle\hat{\Phi}_\parallel|Y\bar{Y}Q_\B|\hat{\Phi}_\parallel\rangle
- \langle\hat{\Phi}_\parallel|J\rangle,\nonumber\\
&= \frac{1}{2}\langle\Phi|\mathcal{P}_{\rm NS}^\dag Y\bar{Y}Q_\B
\mathcal{P}_{\rm NS}|\Phi\rangle
- \langle\Phi|\mathcal{P}_{\rm NS}^\dag|J\rangle.\label{eq:sourceaction}
\end{align}
We can complete the square of this expression by solving the equation of motion
\begin{equation}
\mathcal{P}_{\rm NS}^\dag Y\bar{Y}Q_\B|\hat{\Phi}_\parallel\rangle = \mathcal{P}_{\rm NS}^\dag|J\rangle,
\end{equation}
as
\begin{equation}
|\hat{\Phi}_\parallel\rangle =
\mathcal{P}_{\rm NS}\frac{b_0X_{-\frac{1}{2}}X_{\frac{1}{2}}}{L_0}\mathcal{P}_{Y\bar{Y}}^\dag |J\rangle.
\end{equation}
The action (\ref{eq:sourceaction}) is then rewritten as 
\begin{equation}
S[\Phi, J] = \frac{1}{2}\langle\hat{\Phi}'_\parallel|Y\bar{Y}Q_\B|\hat{\Phi}'_\parallel\rangle
-\frac{1}{2}\langle J|\mathcal{P}_{\rm NS}\frac{b_0X_{-\frac{1}{2}}X_{\frac{1}{2}}}{L_0}\mathcal{P}_{Y\bar{Y}}^\dag
|J\rangle,
\end{equation}
with $|\hat{\Phi}'_\parallel\rangle = |\hat{\Phi}_\parallel\rangle - \Pi_{\rm NS}|J\rangle$.
The propagator $\Pi_{\rm NS}$ is thus found to be
\begin{align}
\Pi_{\rm NS} &= \mathcal{P}_{\rm NS}\frac{b_0X_{-\frac{1}{2}}X_{\frac{1}{2}}}{L_0}\mathcal{P}_{Y\bar{Y}}^\dag
 = \mathcal{P}_{Y\bar{Y}}\frac{b_0X_{-\frac{1}{2}}X_{\frac{1}{2}}}{L_0}\mathcal{P}_{\rm NS}^\dag.
\end{align}
This has quite a reasonable form as a counterpart of the Ramond-Siegel gauge propagator (\ref{def:R-propagator}) in the R sector.

\section{Gauge fixing of total interacting theory}
\label{sec:gf_interacting}

Now, it is straightforward to fix the gauge symmetry of the total interacting theory including both the R and NS sectors with the Yukawa coupling term $\Phi\Psi\Psi$. 
The total action $S$ satisfying the BV master equation is determined by the variational equation
\begin{align}
\delta S &= \int \YY \delta\Phi * \delta_\B\Phi + \int Y \delta\Psi * \delta_\B\Psi,
\end{align}
with the nonlinear BRST transformation
\begin{align}
 \delta_\B \Phi &= Q_\B \Phi 
        + \Phi * \Phi + \bar{X} \Psi * \Psi,\\
 \delta_\B \Psi &= Q_\B \Psi
        + \Phi * \Psi + \Psi * \Phi.
\end{align}
The solution is easily found to be 
\begin{equation}
S= \half\int Y\bar{Y}\,\Phi * \QB \Phi  
+ {1\over3} \int Y\bar{Y}\, \Phi * \Phi * \Phi 
+ \half\int Y\,\Psi * \QB \Psi +  \int Y \Phi * \Psi * \Psi.
\label{eq:masteraction}
\end{equation}
This has the same form as the gauge-invariant action \eqref{eq:action_NS-R}, but now the ghost number constraints on the string fields $\Phi$ and $\Psi$ are relaxed.
As in the case with the free theory, we can restrict the string fields $\Phi$ and $\Psi$ to the hatted ones $\hat{\Phi}=\mathcal{P}_{Y\bar{Y}}\Phi$ and $\hat{\Psi}=\mathcal{P}_Y\Psi$, because of the presence of the inverse picture-changing operators $Y$ and $Y\bar{Y}$ at the interaction
terms and also the property of the NS-R-R vertex 
\begin{align}
 \langle V(1_{\rm NS},2_{\rm R},3_{\rm R})|\Prj_{\YY}^{(1)} &=
\langle V(1_{\rm NS},2_{\rm R},3_{\rm R})|,
\end{align}
resulting from Eq.~\eqref{eq:vprop}.
The gauge-fixed action is finally obtained as
\begin{align}
S &= \half\int Y\bar{Y}\,\hat{\Phi}_\parallel * \QB \hat{\Phi}_\parallel  
+ {1\over3} \int Y\bar{Y}\, \hat{\Phi}_\parallel * \hat{\Phi}_\parallel * \hat{\Phi}_\parallel  
\nn&\qquad\qquad 
+ \half\int Y\,\hat{\Psi}_\parallel* \QB \hat{\Psi}_\parallel
+  \int Y \hat{\Phi}_\parallel * \hat{\Psi}_\parallel* \hat{\Psi}_\parallel\ ,
\label{eq:gfaction}
\end{align}
where $\hat{\Phi}_\parallel$ and $\hat{\Psi}_\parallel$ are the gauge-fixed fields obtained by setting the antifields  equal to zero corresponding to the gauge conditions, that is,
Eqs.~\eqref{eq:gaugecond-NS-1} and \eqref{eq:gaugecond-NS-2} or \eqref{eq:gfcondition} for the NS field
$\hat{\Phi}_\parallel$ and Eq.~\eqref{eq:Ramond-Siegel_condition} for the R field $\hat{\Psi}_\parallel$.

\section{Discussion} 
\label{sec:discussion}

For the NS sector,
we have considered two different gauge-fixing approaches, each of which has
both merits and demerits. In \S \ref{sec:gf-component-NS}, we studied the problem using component fields.
We explicitly chose a set of antifields set equal to zero as gauge conditions:
\begin{align}
& |\psi_{+-}\rangle = |\phi_{+-}\rangle = 0,\\
& |G_\pm\psi_\pm\rangle = 0,\\
& |G_\pm\phi_\pm\rangle - L_+\frac{1}{L_0}|G_\mp\phi_\pm\rangle = 0.
\end{align}
Because the latter two conditions can be written as
\begin{align}
|\psi^*\rangle &= -\frac{1}{L_0}L_+|\psi\rangle,\\
|\phi^*\rangle &= 0,
\end{align}
respectively, using the decomposition \eqref{eq:phipm-decomp} and a similar one for $|\psi_\pm\rangle$,
we can, in principle, compute the BRST transformation of the  {\em fields} $|\phi_0\rangle,\
|\psi_0\rangle,\ |\phi\rangle$, and $|G_\mp\psi_\pm\rangle \propto |\psi\rangle$ from the BRST
transformation (\ref{eq:BRST_NS}).
We could not, however, find the propagator because the explicit form of the kinetic term
is complicated in the component form.

We presented another gauge-fixing approach using the nonlocal projection
operator. We found the gauge condition 
\begin{equation}
 |\hat{\Phi}_\perp\rangle = 
\mathcal{P}_{Y\bar{Y}}\left(1-\frac{b_0}{L_0}Q_\B\right)\mathcal{P}_{Y\bar{Y}}|\hat{\Phi}\rangle = 0,
\end{equation}
which is suitable for computing the propagator.
The propagator in this gauge is given by
\begin{equation}
\Pi_{\rm NS} = \mathcal{P}_{\rm NS}\frac{b_0X_{-\frac{1}{2}}X_{\frac{1}{2}}}{L_0}\mathcal{P}_{Y\bar{Y}}^\dag,
\label{eq:nspropagator}
\end{equation}
in quite a parallel form with the R field propagator
\begin{equation}
 \PiR = \frac{b_0X_0}{L_0}.
\end{equation}
It should also be noted that
the propagator (\ref{eq:nspropagator}) can be rewritten in the form
\begin{equation}
\Pi_{\rm NS} = \mathcal{P}_{Y\bar{Y}}\frac{b_0}{L_0}W
Q_\B\frac{b_0}{L_0}\mathcal{P}_{Y\bar{Y}}^\dag
\end{equation}
with
\begin{equation}
W = \left(X_{-\frac{1}{2}}X_{\frac{1}{2}}Y\bar{Y}X_{-\frac{1}{2}}X_{\frac{1}{2}}\right),
\end{equation}
which is the counterpart of that discussed in Ref.~\citen{Arefeva:1989cp},
although the kernel problem was not considered. 
Because 
\begin{align}
 \Pi_{\rm NS} 
&= \mathcal{P}_{Y\bar{Y}}\left(\frac{b_0 \Xm\Xp}{L_0}-\frac{b_0}{L_0}W\frac{b_0}{L_0}Q_\B\right)\mathcal{P}_{Y\bar{Y}}^\dag 
\nn
&= \mathcal{P}_{Y\bar{Y}}\left(\frac{b_0 \Xm\Xp}{L_0}-Q_\B\frac{b_0}{L_0}W\frac{b_0}{L_0}\right)\mathcal{P}_{Y\bar{Y}}^\dag,
\end{align}
one can see that the physical amplitudes have only the desired pole at $L_0=0$.
This is very favorable in computing perturbative amplitudes.

On the other hand, we cannot determine the BRST transformation of the gauge-fixed field in this approach because $|\hat{\Phi}_\parallel\rangle$ is BRST-invariant:
\begin{equation}
 \delta_\B|\hat{\Phi}_\parallel\rangle = 0.
\end{equation}
This is because $|\hat{\Phi}_\parallel\rangle$ is not the field
itself but also contains the antifield components, as explained in Appendix \ref{app:comp-nonlocal} for the R sector. 
For the NS sector, however, we have not yet identified the field
and antifield components in $|\hat{\Phi}_\parallel\rangle$, because its component-field form is very complicated.

\section*{Acknowledgements}

We would like to thank Masaki Murata for his collaboration in the early stage of this work. 
We also thank Isao Kishimoto for helpful discussions.
This work is 
supported by a Grant-in-Aid for the Global COE Program ``The Next 
Generation of Physics, Spun from Universality and Emergence" from the 
Ministry of Education, Culture, Sports, Science and Technology (MEXT) of
Japan. 
One of the authors (T.K.) is partially supported 
by a Grant-in-Aid for Scientific 
Research (B) (No.\ 20340053) from the Japan Society for the Promotion of
Science (JSPS). 
The work of M.K. is supported by a Grant-in-Aid for JSPS Fellows (No.~21-2291).

\appendix
\section{BRST Operators and BRST Transformations}
\label{app:BRSTop-trf}

\subsection{Ramond string}

We expand the BRST operator for the R string in the ghost zero mode operators $c_0, b_0, \gamma_0$, and $\beta_0$:
\begin{equation}
\QB
= c_0 L_0 + b_0 M + \gamma_0 F + \beta_0 K + \tQB - \gamma_0^2 b_0,
\end{equation}
where 
\begin{subequations}
\begin{align}
L_0 &= L_0^{\rm {(m)}} + \sum_{n\neq0} n:c_{-n}b_n: + \sum_{n\neq0} n:\beta_{-n}\gamma_n:,
\\
M &= -\sum_{n\neq0} n c_{-n}c_{n} - \sum_{n\neq0} \gamma_{-n}\gamma_{n},
\\
F &= F_0^{\rm {(m)}}  - \sum_{n\neq0} \frac12 n \beta_{-n} c_n -2 \sum_{n\neq0} b_{-n} \gamma_{n}, \label{eq:mode-F_R}
\\
K &= \sum_{n\neq0} \frac32 c_{-n} \gamma_{n},
\\
\tQB 
&= \sum_{n\neq0} c_{-n} L_{n}^{{\rm (m)}} 
 + \sum_{n\neq0} \gamma_{-n} F_{n}^{{\rm (m)}}
 + \sum_{\substack{m,n \\ m+n} \neq0} \frac12 (n-m) b_{-n-m}c_n c_m 
\nn&\quad
 + \sum_{\substack{m,n \\ m+n} \neq0} \frac12(2m-n)\beta_{-n-m}c_n\gamma_m 
 - \sum_{\substack{m,n \\ m+n} \neq0} \gamma_{-n}\gamma_{-m} b_{n+m},
\end{align}
\end{subequations}
with $L_n^{\rm (m)}$ and $F_n^{\rm (m)}$ being the super-Virasoro operators for the matter part.
Note that the hermitian conjugate of the BRST operator for the R string is given as\cite{Hata:1987qx}
\begin{equation}
\QB^\dagger = \psi^0_0 \QB \psi^0_0.
\end{equation}

\subsection{Neveu-Schwarz string}

Let us call the ghost and superghost modes $c_0, c_{\pm}, b_0, b_{\pm}, \gamma_{\pm\frac12}$, and $\beta_{\pm\frac12}$ `zero modes' collectively in this NS sector.
We expand the BRST operator for the NS string in these `zero modes':
\begin{align}
\QB 
&= c_0 \left( {\tilde L_0} - c_+b_- - c_-b_+ + \frac12 \gamma_{\frac12}\beta_{-\frac12} -\frac12 \gamma_{-\frac12}\beta_{\frac12} \right)
 + b_0 \left( {\tilde M} + c_-c_+ - 2 \gamma_{-\frac12}\gamma_{\frac12} \right)
\nn
&\quad
 + c_+ {\tilde L}_+ + c_- {\tilde L}_- + {\tilde M}_+ b_+ + {\tilde M}_- b_-
 + \gamma_{-\frac12}{\tilde G}_{\frac12} + \gamma_{\frac12}{\tilde G}_{-\frac12}
 + {\tilde K}_{\frac12} \beta_{-\frac12} + {\tilde K}_{-\frac12} \beta_{\frac12}
\nn
&\quad
 + \frac32 c_{-2} \left( -c_+b_+ - c_-b_+ + c_+b_- + c_-b_- \right)
 + \frac32 c_{2} \left( c_+b_+ - c_-b_+ + c_+b_- - c_-b_- \right)
\nn
&\quad
 + \frac12 c_+ \left( \gamma_{\frac12}\beta_{\frac12} - \gamma_{-\frac12}\beta_{-\frac12} \right)
 - \frac12 c_- \left( \gamma_{\frac12}\beta_{\frac12} + \gamma_{-\frac12}\beta_{-\frac12} \right)
 + c_+ \left( \gamma_{\frac32}\beta_{-\frac12} - \gamma_{-\frac32}\beta_{\frac12} \right)
\nn
&\quad
 - c_- \left( \gamma_{\frac32}\beta_{-\frac12} + \gamma_{-\frac32}\beta_{\frac12} \right)
 - \left( \gamma_{\frac12}^2 + \gamma_{-\frac12}^2 \right) b_+
 + \left( \gamma_{-\frac12}^2 - \gamma_{\frac12}^2 \right) b_-
\nn
&\quad
 - 2 \left( \gamma_{\frac12}\gamma_{-\frac32} + \gamma_{-\frac12}\gamma_{\frac32} \right) b_+
 + 2 \left( \gamma_{\frac12}\gamma_{-\frac32} - \gamma_{-\frac12}\gamma_{\frac32} \right) b_-
 + \tQB.
\end{align}
Here, the tilde symbols on $L_n, M_n, G_r, K_r$, and $\QB$ 
denote the operators with all the parts containing `zero mode' operators omitted:
\begin{subequations}
\begin{align}
{\tilde L}_0
&= L_0^{{\rm (m)}} + \sum_{n\neq0,\pm1} n :b_{-n}c_n: + \sum_{r\neq\pm\frac12} r :\beta_{-r}\gamma_r:,
\\
{\tilde L}_{1}
&= L_{1}^{{\rm (m)}} + \sum_{n\neq0,\pm1,2} (n+1) b_{-n+1}c_n 
 + \sum_{r\neq\pm\frac12,\frac32} \frac12 ( 2r+1 ) \beta_{-r+1} \gamma_r,
\\
{\tilde L}_{-1}
&= L_{-1}^{{\rm (m)}} + \sum_{n\neq0,\pm1,-2} (n-1) b_{-n-1}c_n
 + \sum_{r\neq\pm\frac12,-\frac32} \frac12 ( 2r-1 ) \beta_{-r-1} \gamma_r,
\\
{\tilde L}_{\pm}
&= \frac12 \left( {\tilde L}_{-1} \pm {\tilde L}_1 \right),
\\
{\tilde M}
&= -\sum_{n\neq0,\pm1} n c_{-n}c_{n} - \sum_{r\neq\pm\frac12} \gamma_{-r}\gamma_{r},
\\
{\tilde M}_{1}
&= - \sum_{n\neq0,\pm1,2} \frac12 (2n-1) c_{-n+1} c_n 
 - \sum_{r\neq\pm\frac12,\frac32} \gamma_{-r+1} \gamma_r,
\\
{\tilde M}_{-1}
&= - \sum_{n\neq0,\pm1,-2} \frac12 (2n+1) c_{-n-1} c_n 
 - \sum_{r\neq\pm\frac12,-\frac32} \gamma_{-r-1} \gamma_r,
\\
{\tilde M}_{\pm}
&= {\tilde M}_1 \pm {\tilde M}_{-1},
\\
{\tilde G}_{\pm\frac12}
&= G_{\pm\frac12}^{{\rm (m)}} - \sum_{n\neq0,\pm1} \left( \frac12(n\pm1) \beta_{-n\pm\frac12} c_n + b_{-n}\gamma_{n\pm\frac12} \right),
\\
{\tilde K}_{\pm\frac12}
&= \sum_{n\neq0,\pm1} \frac12(-3n \pm1) c_n \gamma_{-n\pm\frac12},
\\
\tQB 
&= \sum_{n\neq0,\pm1} c_{-n} L_{n}^{{\rm (m)}} 
 + \sum_{r\neq\pm\frac12} \gamma_{-r} G_{r}^{{\rm (m)}}
 + \sum_{\substack{m,n \\ m+n} \neq0,\pm1} \frac12 (n-m) b_{-n-m}c_n c_m
\nn&\quad
 + \sum_{n\neq0,\pm1}\sum_{r\neq\pm\frac12} \left( \frac12(2r-n)\beta_{-n-r}c_n\gamma_r - b_{-n}\gamma_{n-r}\gamma_r \right).
\end{align}
\end{subequations}

The translated BRST operator by $T$ in Eq.~\eqref{def:translationT} can be obtained as
\begin{align}
\Exp{T} \QB \Exp{-T}  \label{eq:hatQB}
 &= \tilde{\mathcal Q}
 + \frac12 \gamma_{\frac12} (\dd C_+ - C_-) \beta_{-\frac12} 
 + \frac12 \gamma_{-\frac12} (\dd C_+ - C_-)   \beta_{\frac12} 
\nn&\quad
 + \frac12 \gamma_{\frac12} (\dd C_- + C_+) \beta_{\frac12} 
 - \frac12 \gamma_{-\frac12} (\dd C_- + C_+) \beta_{-\frac12} 
 + \gamma_{\frac12} {\tilde G}_{-\frac12}  
 + \gamma_{-\frac12} {\tilde G}_{\frac12} 
\nn&\quad
 + \tilde{\mathcal M} b_+ 
 + 2 \gamma_{\frac12} \left(\Gamma_+ - \gamma_{-\frac32}\right) b_+ 
 + 2 \gamma_{-\frac12} \left(\Gamma_- - \gamma_{\frac32}\right) b_+ 
 - \left( \gamma_{\frac12}^2 + \gamma_{-\frac12}^2 \right) b_+
\nn&\quad
 + \frac32 c_+ (c_{-2} - c_{2}) b_+
 + c_+ {\tilde L}_+ 
 + \frac12 c_+ \gamma_{\frac12} \beta_{\frac12}
 - \frac12 c_+ \gamma_{-\frac12} \beta_{-\frac12},
\end{align}
with
\begin{align}
\tilde{\mathcal Q} 
 &= \tQB - C_-{\tilde L}_0 - C_+{\tilde L}_- - \Gamma_-{\tilde G}_{\frac12} - \Gamma_+{\tilde G}_{-\frac12}, 
\\
\tilde{\mathcal M} 
 &= {\tilde M}_+ - C_-C_+ + \frac32 (c_{-2} + c_2) C_+ - \Gamma_+^2 - \Gamma_-^2 + 2\Gamma_- \gamma_{\frac32} + 2 \Gamma_+ \gamma_{-\frac32},
\\
\dd c _+(i) &= \frac12\big( \dd c(i) + \dd c(-i) \big)
 = c_0 + \dd C_+,
\\
\dd C_+ &= \sum_{k=1}^\infty (-1)^k \big( (-2k+1) c_{2k} +(2k+1)c_{-2k} \big),
\\
\dd c _-(i) &= \frac{1}{2i}\big( \dd c(i) - \dd c(-i) \big)
 = 2 c_{-1} + \dd C_-,
\\
\dd C_- &= i \sum_{k=1}^\infty (-1)^k \big( 2k c_{2k+1} +2(k+1)c_{-2k-1} \big).
\end{align}
Note that in obtaining Eq.~\eqref{eq:hatQB} we have eliminated the terms containing $c_0$ and $c_-$ on the left or $b_0$ and $b_-$ on the right.
This is because we use it only when the $(\YY)_0$ operator is present on the left and the reduced field  $\Exp{T} \Prj_0 \NS$ on the right.

The expression of the BRST transformation in component fields can be most easily obtained by computing 
\begin{equation}
\Exp{T} \Big( \delB \tilde \NS \Big) = \Exp{T} \Prj_0 \Exp{-T} \cdot \Exp{T} \QB \Exp{-T} \cdot \Exp{T} \Prj_0 \NS,
\end{equation}
using Eq.~\eqref{eq:hatQB}. The result is
\begin{subequations} \label{eq:BRST_NS}
\begin{align}
\delB \phi_0 
&= \tilde{\mathcal Q} \phi_0 
 + \tilde{\mathcal M} \psi_0,
\\
\delB \phi_{\pm} 
&= {\tilde G}_{\mp\frac12} \phi_0 
 + \left( \tilde{\mathcal Q} - \frac12 (\dd C_+ - C_-)  \right) \phi_{\pm}
\mp \frac12 (\dd C_- + C_+) \phi_{\mp}
\nn&\quad
+ 2\left(\Gamma_+ - \gamma_{-\frac32} \right) \psi_0
+ \tilde{\mathcal M} \psi_{\pm},
\\
\delB \phi_{+-} 
&= {\tilde G}_{-\frac12} \phi_{+} + {\tilde G}_{\frac12} \phi_{-} 
 + \left( \tilde{\mathcal Q} - \dd C_+ + C_- \right) \phi_{+-}
\nn&\quad
 - 2 \psi_0 
 + 2\left(\Gamma_+ - \gamma_{-\frac32} \right) \psi_{+}
 + 2\left(\Gamma_- - \gamma_{\frac32} \right) \psi_{-}
 + \tilde{\mathcal M} \psi_{+-},
\\
\delB \psi_0 
&= - {\tilde L}_+ \phi_0
 + \left( \tilde{\mathcal Q}  + \frac32 (c_2-c_{-2}) \right) \psi_0,
\\
\delB\psi_{\pm} 
&= - {\tilde L}_+ \phi_{\pm} \pm \frac12 \phi_{\mp} 
 + {\tilde G}_{\mp\frac12} \psi_0 
 \mp \frac12 (\dd C_- + C_+) \psi_{\mp}
\nn&\quad
 + \left( \tilde{\mathcal Q} - \frac12 (\dd C_+ - C_-) + \frac32 (c_2-c_{-2}) \right) \psi_{\pm},
\\
\delB\psi_{+-}
&= - {\tilde L}_+ \phi_{+-} 
 + {\tilde G}_{-\frac12} \psi_{+}
 + {\tilde G}_{\frac12} \psi_{-}
\nn&\quad
 + \left( \tilde{\mathcal Q} - \dd C_+ + C_- + \frac32 (c_2-c_{-2}) \right) \psi_{+-}.
\end{align}
\end{subequations}

\section{Picture Changing Operators} 
\label{app:PCO}

The picture-changing operator $X(z)$ and the inverse picture-changing operator $Y(z)$ are defined as
\begin{align} 
X(z) &= [\QB, \Theta(\beta(z))]
= G(z)\delta(\beta(z)) - \dd b \delta'(\beta(z)), \label{def:PCO_Xz}
\\
Y(z) &= c(z)\delta'(\gamma(z)), \label{def:PCO_Yz}
\end{align}
where $G(z)$ denotes the supercurrent
\begin{equation}
G(z) = G^{\rm (m)}(z) + c\dd\beta(z) + \frac32 (\dd c) \beta (z)- 2 \gamma b (z).
\end{equation}
The operators $X(z)$ and $Y(z)$ are the inverses of each other
and change the picture numbers by $+1$ and $-1$, respectively.
We write in particular the (inverse) picture-changing operators inserted at the midpoint or its mirror point $z=\pm\ii$ as
\begin{equation}
Y = Y(\ii), \quad \bar Y = Y(-\ii),\quad X=X(\ii),\quad \bar X =X(-\ii).
\end{equation}

The `mode versions' of the (inverse) picture-changing operators are defined by
\begin{align} 
X_0 &= [\QB, \Theta(\beta_0) ] 
 = \delta(\beta_0)F_0 - b_0 \delta'(\beta_0), \label{def:PCO_X0}
\\
X_{\pm\frac{1}{2}} 
&= [\QB, \Theta(\beta_{\pm\frac12})] 
 = \delta(\beta_{\pm\frac12}) G_{\pm\frac12} - b_{\pm1} \delta'(\beta_{\pm\frac12}), \label{def:PCO_Xpm}
\end{align}
and 
\begin{align}
Y_0 &= c_0\delta'(\gamma_0), \label{def:PCO_Y0}
\\
(Y\bar Y)_0 &= \frac14 c_- c_0 \left( \delta''(\gamma_{\frac12})\delta(\gamma_{-\frac12}) + \delta(\gamma_{\frac12})\delta''(\gamma_{-\frac12}) \right). \label{def:PCO_YYb0}
\end{align}
They satisfy the relations:
\begin{subequations} \label{eq:XYX_Ramond}
\begin{alignat}{2} 
Y(z)X_0Y(z) &= Y(z), &\qquad X_0 Y(z) X_0 &= X_0, 
\\
Y_0X_0Y_0 &= Y_0, &\qquad X_0 Y_0 X_0 &= X_0,
\end{alignat}
\end{subequations}
and 
\begin{equation}
Y\bar Y(z)\Xm\Xp Y\bar Y(z) = Y\bar Y(z).
\end{equation}
In addition, $X_0$ and $\Xm\Xp$ (anti-)commute with $b_0$ and $L_0$:
\begin{alignat}{2}
\{X_0, b_0\} &= 0,
&\qquad
 [X_0, L_0] &= 0,
\\
\{X_{\pm\frac12}, b_0\} &= 0,
&\qquad
[X_{\pm\frac12}, L_0] &= \pm\frac12 X_{\pm\frac12}.
\end{alignat}

The hermitian conjugates of the (inverse) picture-changing operators are given as
\begin{align}
Y(\pm\ii)^\dagger &= Y(\pm\ii),
\\
X_0^\dagger &= \psi^0_0 X_0 \psi^0_0,  \label{eq:X_0dagger}
\\
(\Xm \Xp)^\dagger &= \Xm \Xp,
\end{align}
where $\psi^{\mu=0}_0$ is the zero mode of the time component of the matter operator $\psi^{\mu}(z)$ in the R sector.
Accordingly, we have to distinguish the hermitian conjugate and the transposition of the Ramond projection $\Prj_Y$ given as
\begin{equation}
\Prj_Y^\dagger = Y X_0^\dagger = Y \psi^0_0 X_0 \psi^0_0,
\qquad
\Prj_Y^T = Y X_0,
\end{equation}
and can confirm the relations
\begin{equation}
\psi^0_0 \Prj_Y^\dagger = \Prj_Y^T \psi^0_0,
\qquad
\Prj_Y^\dagger \psi^0_0 = \psi^0_0 \Prj_Y^T.
\end{equation}

\section[Ramond analog of Decomposition $(6.6)$]{Ramond Analog of Decomposition \eqref{eq:decompA}}
\label{app:comp-nonlocal}

In this appendix, we explain how the decomposition using the nonlocal 
projection operator gives the proper gauge condition and the propagator 
for the R string, which can be explicitly studied by component analysis. 
The nonlocal projection operator of the R 
string has the form
\begin{align}
{\mathcal{P}}_b &= \frac{b_0}{L_0}Q_\B,\\
&= b_0c_0 -\frac{1}{L_0}(F\gamma_0+K\beta_0+\tQB)b_0.
\end{align}
Unlike the NS string, this satisfies
\begin{equation}
 \mathcal{P}_Y{\mathcal{P}}_b\mathcal{P}_Y = {\mathcal{P}}_b\mathcal{P}_Y. 
\end{equation}
The decomposition of the projected field $|\hat{\Psi}\rangle=\mathcal{P}_Y|\Psi\rangle$ is defined by
\begin{align}
 |\hat{\Psi}\rangle &= \mathcal{P}_b|\hat{\Psi}\rangle + (1-\mathcal{P}_b)|\hat{\Psi}\rangle,\\
&\equiv |\hat{\Psi}_\parallel\rangle + |\hat{\Psi}_\perp\rangle,
\end{align}
with which the gauge transformation $ \delta|\hat{\Psi}\rangle = Q_\B|\hat{\Lambda}\rangle$ splits into
\begin{align}
 \delta|\hat{\Psi}_\parallel\rangle &= 0,\\
 \delta|\hat{\Psi}_\perp\rangle &= Q_\B|\Lambda_\parallel\rangle.
\end{align}
Therefore, we can gauge away the $|\hat{\Psi}_\perp\rangle$ part and take the gauge condition
\begin{equation}
 |\hat{\Psi}_\perp\rangle = 0.\label{eq:rgcondition}
\end{equation}

As explained in \S \ref{sec:gaugefix-R}, the constrained string field
$|\hat{\Psi}\rangle$ can be expressed by the component field as
\begin{equation}
 |\hat{\Psi}\rangle = |\phi\rangle - (\gamma_0+c_0F)|\psi\rangle.
\end{equation}
The decomposed fields $|\hat{\Psi}_\parallel\rangle$ and $|\hat{\Psi}_\perp\rangle$ are
also represented by $|\phi\rangle$ and $|\psi\rangle$ as
\begin{align}
 |\hat{\Psi}_\parallel\rangle &= |\phi\rangle + \tQB\frac{F}{L_0}|\psi\rangle,\label{eq:parallel}\\
 |\hat{\Psi}_\perp\rangle &= - \left(\gamma_0+c_0F+\tQB\frac{F}{L_0}\right)|\psi\rangle. \label{eq:Rperp}
\end{align}
Since this \eqref{eq:Rperp} also leads to $\ket{\psi}=\beta_0\fket{\hat\Psi_\perp}$, the gauge condition (\ref{eq:rgcondition}) is equivalent to the gauge condition $|\psi\rangle = 0$ used in \S \ref{sec:gaugefix-R}.

We can also obtain the same propagator as that in \S \ref{sec:gaugefix-R} using this argument; indeed,
\begin{align}
 S &= -\frac{1}{2}\langle\hat{\Psi}_\parallel|Y_0 Q_\B|\hat{\Psi}_\parallel\rangle 
+ \langle\hat{\Psi}_\parallel|J\rangle,\\
&= \frac{1}{2}\left(\langle J|\frac{X_0b_0}{L_0}-\langle\hat{\Psi}_\parallel|\right)Y_0 Q_\B
\left(|\hat{\Psi}_\parallel\rangle-\frac{b_0X_0}{L_0}|J\rangle\right)
-\frac{1}{2}\langle J|\frac{b_0X_0}{L_0}|J\rangle.
\end{align}

We can see that the component $|\hat{\Psi}_\parallel\rangle$ (\ref{eq:parallel}) contains not only the field $\ket{\phi}$ but also the antifield $|\psi\rangle$.The proper BRST transformation of the gauge-fixed field can be obtained only when the antifield component is set equal to zero before the transformation:
\begin{equation}
 \delta_\B\left(|\hat{\Psi}_\parallel\rangle|_{|\psi\rangle=0}\right) 
= \tQB\left(|\hat{\Psi}_\parallel\rangle|_{|\psi\rangle=0}\right).
\end{equation}
Otherwise, the BRST transformation of $\fket{\PPRamond}$ is zero.



\begin{thebibliography}{99}

\bibitem{Witten:1986qs}
  E.~Witten,
  \NPB{276,1986,291}.
\bibitem{Witten:1985cc}
  E.~Witten,
  \NPB{268,1986,253}.
\bibitem{Wendt:1987zh}
  C.~Wendt,
  \NPB{314,1989,209}.
\bibitem{Preitschopf:1989fc}
  C.~R.~Preitschopf, C.~B.~Thorn and S.~A.~Yost,
  \NPB{337,1990,363}.
\bibitem{Arefeva:1989cp}
  I.~Y.~Arefeva, P.~B.~Medvedev and A.~P.~Zubarev,
  \NPB{341,1990,464}.
\bibitem{Berkovits:1995ab}
  N.~Berkovits,
  \NPB{450,1995,90}
  [Errata,\  B {\bf 459} (1996), 439]
  hep-th/9503099.
\bibitem{Michishita:2004by}
  Y.~Michishita,
  \JHEP{01,2005,012},
  hep-th/0412215.

\bibitem{Kugo:2009APCTP}
  M.~Kohriki, T.~Kugo, H.~Kunitomo and M.~Murata,
  Talk given by T.~Kugo at APCTP 
  ``APCTP Focus Program on Current Trends in String Field Theory"
  http://newton.skku.ac.kr/workshop/SFT2009/program.html
\bibitem{Kohriki:2011zz}
  I.~Kishimoto, M.~Kohriki, T.~Kugo, H.~Kunitomo and M.~Murata,
  \PTPS{188,2011,263}.

\bibitem{Kugo:1988mf}
  T.~Kugo and H.~Terao,
  \PLB{208,1988,416}.
\bibitem{Arefeva:1988nn}
  I.~Y.~Arefeva and P.~B.~Medvedev,
  \PLB{212,1988,299}.
\bibitem{Kazama:1986cy}
  Y.~Kazama, A.~Neveu, H.~Nicolai and P.~C.~West,
  \NPB{278,1986,833}.

\bibitem{Terao:1986ex}
  H.~Terao and S.~Uehara,
  \PLB{173,1986,134}.
\bibitem{Hata:1986zq}
  H.~Hata, K.~Itoh, T.~Kugo, H.~Kunitomo and K.~Ogawa,
  \NPB{283,1987,433}.
\bibitem{Banks:1985xa}
  T.~Banks, M.~E.~Peskin, C.~R.~Preitschopf, D.~Friedan and E.~J.~Martinec,
  \NPB{274,1986,71}.
\bibitem{Kiermaier:2007jg}
  M.~Kiermaier, A.~Sen and B.~Zwiebach,
  \JHEP{03,2008,050},
  arXiv:0712.0627.
\bibitem{Thorn:1986qj}
  C.~B.~Thorn,
  \NPB{287,1987,61}.
\bibitem{Bochicchio:1986}
  M.~Bochicchio,
  \PLB{188,1987,330}
  ;
  \PLB{193,1987,31}.

\bibitem{Urosevic:1990as}
  B.~V.~Urosevic and A.~P.~Zubarev,
  \PLB{246,1990,391}.

\bibitem{Hata:1987qx}
  H.~Hata, K.~Itoh, T.~Kugo, H.~Kunitomo and K.~Ogawa,
  \PTP{78,1987,453}.

\end{thebibliography}
\end{document}